\documentclass[10pt,aps,twocolumn,pra,floatfix,superscriptaddress]{revtex4-2}

\usepackage{xcolor}

\usepackage{graphicx}
\usepackage{dcolumn}
\usepackage{bm}

\usepackage{amsfonts}
\usepackage{xurl} 
\usepackage[breaklinks=true, hidelinks]{hyperref}
\usepackage{amssymb}
\usepackage{amsmath}
\usepackage{calc}
\usepackage{graphicx}
\usepackage{bm}
\usepackage{verbatim}

\def\be{ \begin{equation} }
\def\ee{ \end{equation} }
\def\bea{ \begin{align} }
\def\eea{ \end{align} }
\def\bse{ \begin{subequations} }
\def\ese{ \end{subequations} }

\usepackage{xcolor}

\begin{document}

\title{Composite quantum gates simultaneously compensated for multiple errors}

\author{Hristo G. Tonchev}
\affiliation{Center for Quantum Technologies, Department of Physics, Sofia University, 5 James Bourchier Boulevard, 1164 Sofia, Bulgaria}
\affiliation{Georgi Nadjakov Institute of Solid State Physics, Bulgarian Academy of Sciences, 72 Tsarigradsko Chaussée, 1784 Sofia, Bulgaria}

\author{Nikolay V. Vitanov}
\affiliation{Center for Quantum Technologies, Department of Physics, Sofia University, 5 James Bourchier Boulevard, 1164 Sofia, Bulgaria}

\date{October 23, 2026}

\begin{abstract}
Systematic control errors remain a primary obstacle to realizing high-fidelity single-qubit gates. To suppress these errors, we construct phase-controlled and variable-area $X$ and Hadamard composite gates that compensate for simultaneous quasistatic amplitude and detuning errors in a two-level rectangular-pulse model. Our construction uses two complementary strategies: (i) cancellation of error derivatives of the full unitary propagator in the Cayley–Klein parametrization, and (ii) direct minimization of the average gate infidelity over prescribed error ranges. For shorter sequences, our approach yields performance comparable to established analytical solutions. Selected longer sequences achieve substantially lower mean
gate infidelities than established comparison pulses over
the prescribed error domains, including reductions of
several orders of magnitude. Across the derived composite-gate families, the central mean infidelity decreases approximately exponentially with sequence length. Selected controls retain high fidelity in simulations with finite pulse edges and time-dependent detuning, without a need for recalibration.

\end{abstract}

\maketitle

\section{Introduction}

Composite pulse sequences (CPs) \cite{PhysRevLett.113.043001,arbitraryCp,Brown_2004,Transmon_CP,counsell1985analytical} offer a powerful quantum control method \cite{qc1,qc2,dong2010quantum,d2021introduction,chernev2026universalcompositephasegates} for suppressing systematic errors in the parameters that describe the interaction of a quantum system with an external electromagnetic field.
A composite pulse is a sequence of (identical or non-identical) pulses with well-defined relative phases between them, which are used as control parameters.

Originally, they were developed in the field of nuclear magnetic resonance in the early 1980s \cite{FREEMAN1980453,LEVITT198661} with the objective of expanding the high-excitation range as much as possible (broadband CPs), or, alternatively, to shrink it as much as possible \cite{NRPBCP,BoyqnNB} (narrowband CPs). 
In the former case, robust excitation is achieved in a broad range of parameter variations, whereas in the latter case, greater selectivity is delivered.

In the early decades, the main performance benchmark was the bandwidth of the pulse.
With the development of quantum technologies, increasing attention has been directed toward accurate implementation of quantum gates, requiring control of both transition probabilities and relative phases \cite{Shi_2022,fidelityCp,wolfibook,qgates1,qgates2,kajsagates}.
CPs have proved to be a leading contender in this respect as, in addition to robust (or sensitive) excitation probability, they offer extremely low errors \cite{torosov2011high}. They have also been applied to liquid-state NMR~\cite{Bando_2020_NMR}, trapped ions~\cite{Timoney_2008}, and superconducting circuits~\cite{Collin_2004}. 

CPs can be divided into two major groups: variable and constant rotations \cite{SAKELLARIOU200225}.
For variable rotations, only the transition probability is compensated for errors, but the phases of the propagator remain sensitive to errors.
For constant rotations, both the probability and the phases are compensated for parameter errors, and so is the entire propagator.
In order to construct quantum gates, constant rotations are the obvious correct choice.
However, they generally impose more constraints than compensation of the transition probability alone.

Established constant-rotation sequences include BB1~\cite{NRPBCP} and SK1~\cite{Brown_2004} and SCROFULOUS~\cite{Cummins_2003} for amplitude errors, and CORPSE~\cite{Cummins_2003} for detuning errors. Simultaneous correction is provided by concatenated constructions, including reduced CORPSE-in-SK1 and CORPSE-in-BB1 for arbitrary transverse rotations~\cite{Bando_2013}, planar-geometric and switchback constructions~\cite{Ichikawa_2014_planar,Kukita_2022_switchback}, and short analytical NOT gates~\cite{Jones_2013_NOT,Odedra_2012}. SUPCODE also addresses simultaneous quasistatic errors with the different controls of singlet--triplet qubits~\cite{Wang_2014_SUPCODE}. Universal CPs provide a broad population-inversion guarantee under identical constituent-pulse errors~\cite{PhysRevLett.113.043001,Genov_2020_profile}, but this guarantee alone does not ensure the relative phase required for a specified quantum gate. We examine this distinction in Sec.~\ref{Sec:X gates}.

Here we construct composite $X$ and $R_x(\pi/2)$ gates for coherent quasistatic amplitude and detuning errors in a rectangular-pulse model, with ideal virtual \(Z\) rotations completing the Hadamard gate. The five-pulse derivative solutions recover known universal phase families after axis alignment, and the
seven-pulse derivative solutions select particular members of Jones's symmetric family~\cite{Jones_2013_NOT}. We extend the construction to longer phase-controlled and variable-area sequences, exploring how the additional control freedom shapes gate accuracy over finite ranges of amplitude and detuning errors. The resulting families are studied alongside established analytical and concatenated composite pulses to examine the relation between their sequence length, pulse area, and robustness.

For variable-area designs containing up to 15 pulses, we distinguish two objectives: deeper suppression within a prescribed calibration window and broader tolerance at a fixed infidelity threshold.  Under paired pulse-count and area limits, this choice substantially changes the connected high-fidelity region. Local derivative cancellation, finite-domain mean error, and threshold width therefore characterize different aspects of robustness. Our pulses also address duration errors through an exact parameter mapping, and phase-precision and finite-edge tests quantify limitations of ideal control. A similar approach was previously used to create $X$ and $\sqrt{X}$ CP
gates in a multilevel transmon model~\cite{Transmon_CP}.

Section~\ref{Sec:derivation} defines the model and design methods. Sections~\ref{Sec:X gates} and \ref{Sec:Hadamard} present the $X$ and Hadamard constructions and their reference comparisons. Section~\ref{Sec:objective_selection} analyzes pulse-area costs and the trade-off between depth and width. The appendices give the control specifications and supporting phase-precision and finite-edge tests.

\section{Derivation \label{Sec:derivation}}

The propagator describing coherent interaction of a qubit with an external field can be parameterized with two complex-valued Cayley-Klein parameters \cite{Shore_2011} $a$ and $b$ (with $|a|^2+|b|^2=1$) as
\be
\mathbf{U} = 
\left[\begin{array}{cc}
a & b \\
-b^\ast & a^\ast
\end{array}
\right] .
\label{eq:kleinparms}
\ee
The transition probability is the squared modulus of the off-diagonal element, $P=|b|^2$. 

For the framework developed in this paper, the propagator in Eq.~\eqref{eq:kleinparms} is generated by the standard RWA single-qubit Hamiltonian. We assume exact resonance at the nominal operating point, $\Delta_0=0$, and introduce a fixed positive Rabi frequency $\Omega_0$. We define an amplitude error $\epsilon$ by $\Omega=\Omega_0(1+\epsilon)$, whereas the dimensionless detuning error is $\delta=\Delta/\Omega_0$, so that $\Delta=\delta\Omega_0$. Thus $\delta$ describes an additive frequency offset from resonance, normalized by the nominal Rabi frequency. Setting $\hbar=1$ and initially taking the drive phase to be zero, the Hamiltonian is
\begin{equation}
\mathbf{H} = \frac{\Omega_0}{2} \left[\begin{array}{cc} -\delta & 1+\epsilon \\ 1+\epsilon & \delta \end{array} \right].
\label{eq:Hamiltonian}
\end{equation}
The reference $\Omega_0$ is independent of both error parameters and the detuning is not rescaled by the amplitude $\Omega_0(1+\epsilon)$. The nominal operating point is $(\epsilon,\delta)=(0,0)$. For variable-area sequences, the actual amplitude of pulse $k$ is $\Omega_k(1+\epsilon)$, while the common detuning remains $\Delta=\delta\Omega_0$, independent of $k$.
The analytic derivations and rectangular-pulse benchmarks assume constant detuning. For a pulse of duration $T$, the propagator is $\mathbf{U}=\exp(-i\mathbf{H}T)$. Throughout the manuscript we use the common reference $\Omega_0T=\pi$. A single $X$ pulse has amplitude $\Omega_0$, whereas a single $R_x(\pi/2)$ primitive has amplitude $\Omega_0/2$ at the same duration $T$.
Appendix~\ref{app:pulse_shape} tests selected controls with short smooth edges and continuous time-dependent detuning.

From Eqs.~\eqref{eq:kleinparms} and \eqref{eq:Hamiltonian}, it follows that introducing a constant phase shift to the drive field, $\Omega \to \Omega e^{i\phi}$, maps onto the propagator as a straightforward transformation:
\begin{equation}
\mathbf{U}(\phi) = \left[\begin{array}{cc} a & b e^{i\phi} \\ -b^\ast e^{-i\phi} & a^\ast \end{array} \right] .
\end{equation}

The CP sequence is then described by a sequence of $N$ such propagators (where the subscript $k$ denotes pulse-specific parameters),
\begin{equation}
\mathbf{U}_k(\phi_k) = \left[\begin{array}{cc} a_k & b_k e^{i\phi_k} \\ -b_k^\ast e^{-i\phi_k} & a_k^\ast \end{array} \right] ,
\end{equation}
and the overall sequence produces the total effective propagator
\begin{equation}
\mathcal{U} = \mathbf{U}_N(\phi_N) \cdots \mathbf{U}_2(\phi_2) \mathbf{U}_1(\phi_1) .
\end{equation}

One way to implement the control phases on a physical system is as digital updates of the drive phase, using the same mechanism as virtual $Z$ rotations \cite{McKay_2017}. Finite digital phase resolution and residual physical phase errors are considered separately in Appendix~\ref{app:phase_precision}.

Symmetric $X$ designs use identical nominal $\pi$ pulses with adjustable phases; numerical $X$ and Hadamard designs also vary the nominal amplitudes $\Omega_k$ at equal segment duration. The common errors $\epsilon$ and $\delta$ are constant during a sequence, and the detuning remains $\Delta=\delta\Omega_0$ for every pulse.

We use two design strategies for suppressing both errors.

Our \textit{first approach}, which follows the conventional strategy, is to enforce that the propagator $\mathcal{U}$ matches the desired quantum gate $\mathcal{G}$ at the nominal values,
\be
\mathcal{U}(0,0) = \mathcal{G},
\ee
and to cancel as many error derivatives as possible, i.e.
\be
D_{m,n}\mathcal{U}_{jk} = \frac{\partial^{m+n}}{\partial \epsilon^m \partial \delta^n} \, \mathcal{U}_{jk} = 0,
\quad \text{for } m+n \geq 1,
\label{eq:derivatives}
\ee
up to the desired order, with all derivatives evaluated at $(\epsilon,\delta)=(0,0)$.
If every derivative of total order $1\leq m+n\leq p$ vanishes for an exactly matched target, then $\mathcal U-\mathcal G=\mathcal O(r^{p+1})$ and the average gate infidelity is $\mathcal O(r^{2p+2})$, where $r=\sqrt{\epsilon^2+\delta^2}$.

In our \textit{second approach}, we minimize the mean gate infidelity over a discrete grid of amplitude and detuning errors. At each error pair $(\epsilon,\delta)$, the infidelity is
\begin{equation}
\mathcal I(\epsilon,\delta)
=1-F\!\left(U(\epsilon,\delta),\mathcal G\right),
\label{eq:infidelity}
\end{equation}
where $F$ is the average gate fidelity between the implemented propagator and the target gate~\cite{Nielsen_2002,Pedersen_2007},
\begin{equation}
F(U,\mathcal G)
=\frac{|\mathrm{Tr}(U^\dagger\mathcal G)|^2+d}{d(d+1)}.
\label{eq:fidelity_function}
\end{equation}
Here $d$ is the Hilbert-space dimension ($d=2$ for a qubit).

To average over control errors, we sample $N_\epsilon$ amplitude values $\epsilon_j\in[\epsilon_{\min},\epsilon_{\max}]$ and $N_\delta$ detuning values $\delta_k\in[\delta_{\min},\delta_{\max}]$. The equally weighted grid objective is
\begin{equation}
\overline{\mathcal I}_{\rm grid}
=\frac{1}{N_\epsilon N_\delta}
\sum_{j=1}^{N_\epsilon}\sum_{k=1}^{N_\delta}
\mathcal I(\epsilon_j,\delta_k).
\label{eq:grid_infidelity}
\end{equation}
This mean objective rewards accuracy throughout the chosen error window. For wider designs, where we want to increase the accurate region without compromising the area around the nominal point we modify Eq.~\eqref{eq:grid_infidelity} and instead use
\begin{equation}
J_w(Q)=\frac1M\sum_{q\in Q}
\frac{\mathcal I(q)}{\mathcal I(q)+\tau},
\qquad \tau=10^{-4},
\label{eq:width_loss}
\end{equation}
where $Q$ contains $M$ equally weighted error samples from a wider region. Points with infidelity far below $\tau$ contribute little to this objective, so the search favors a larger accurate region over further suppression of a small central error. The error windows, pulse-area bounds and peak-amplitude bounds are specified together with the corresponding sequences.

The Hadamard searches apply these same two objectives to rectangular pulses and to a fixed finite-edge model, discussed in Appendix~\ref{app:pulse_shape}. For precision designs, the mean objective is
\begin{equation}
J_c=0.3\,\overline{\mathcal I}_{\mathrm{rect},0.15}
+0.7\,\overline{\mathcal I}_{\mathrm{edge},0.15},
\label{eq:H_selection}
\end{equation}
where the subscript specifies the error-window half-width. The coverage search combines the rectangular and finite-edge evaluations of $J_w$ with the same weights. The edges are area-preserving cosine ramps of 4 ns at a common peak rate $\Omega_{\max}/(2\pi)=5$ MHz. 
The finite-edge model and its numerical validation are
described in Appendix~\ref{app:pulse_shape}. Detailed
optimization settings and candidate-selection records
are provided in the reproduction
materials~\cite{my_github_repo}.

The same design problem can be formulated for other rotation angles by choosing the target
\begin{equation}
\mathcal{G}_\theta=R_x(\theta)=\exp(-i\theta\sigma_x/2),
\label{eq:general_target}
\end{equation}
where $\sigma_x$ is the first Pauli matrix.
Together with virtual $Z$ rotations, $Z(\alpha)R_x(\theta)Z(\beta)$ this covers arbitrary single-qubit unitaries up to global phase. Here we solve and benchmark the cases only for $\theta=\pi$ and $\pi/2$, but expect that this approach can be extended.

\section{X gates \label{Sec:X gates}}

We represent the $X$ gate up to a global phase as,
\be\label{targetX}
\mathcal{G} = -iX = e^{-i\pi\sigma_x/2}.
\ee
For this gate, we consider symmetric sequences of nominal $\pi$ pulses and numerical variable-area sequences without symmetry constraint.
The symmetric sequences contain $2n-1$ pulses with the anagram phases,
\be
\pi_{\phi_1} \pi_{\phi_2} \cdots \pi_{\phi_n} \cdots \pi_{\phi_2} \pi_{\phi_1} .
\ee
Palindromic controls allow for easier derivations, yielding analytic expressions for the composite phases of relatively short pulse sequences. Moreover, the fidelity landscape is symmetric with respect to detuning errors. In contrast, asymmetric sequences enable higher-order error correction with the same number of pulses, but produce an asymmetric fidelity landscape.

\subsection{Composite sequences of three pulses}

\begin{figure}
\includegraphics[width=0.99\columnwidth]{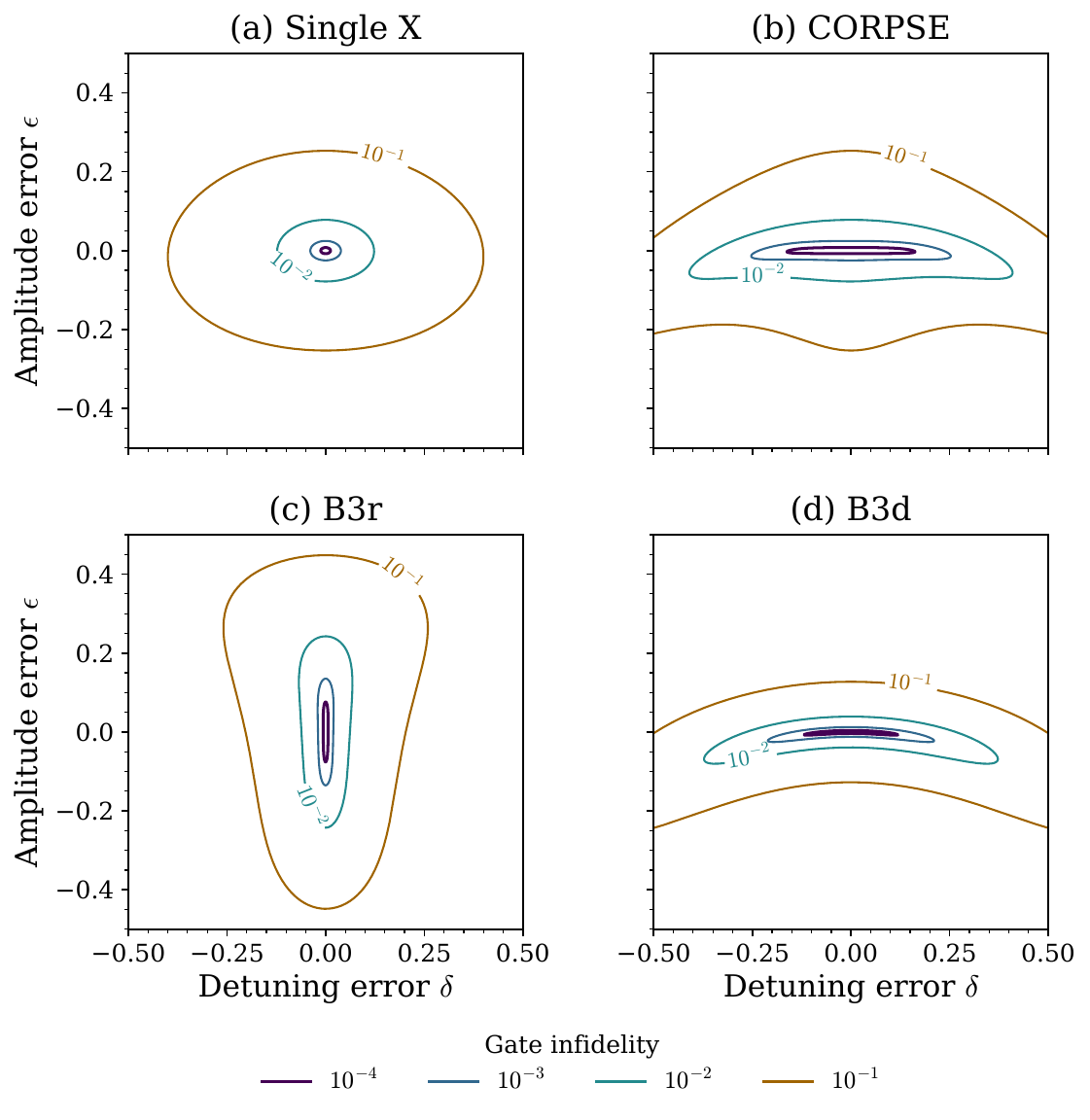}

    \caption{(Color online)
Infidelity of X gate versus the detuning error $\delta=\Delta/\Omega_0$ and the Rabi frequency error $\epsilon$ for (a) single $\pi$ pulse, (b) CORPSE, (c) B3r pulse, (d) B3d pulse.
Contours show $10^{-4}$ (emphasized purple), $10^{-3}$ (blue), $10^{-2}$ (teal), and $10^{-1}$ (ochre). Subsequent landscape figures retain these colors and include finer contours where specified.
    }
    \label{fig:X3}
\end{figure}

We benchmark three well-known three-pulse sequences, using the notation $(\theta)_\phi$ for a pulse of area $\theta$ and phase $\phi$. The first is the CORPSE sequence \cite{TYCKO198590,Cummins_2003},
\begin{equation}
\mathrm{CORPSE}:\; \left(\tfrac{7\pi}{3}\right)_0\; \left(\tfrac{5\pi}{3}\right)_\pi\; \left(\tfrac{\pi}{3}\right)_0 .
\end{equation}
For CORPSE, all three pulses have the same nominal Rabi frequency $\Omega_0$ and their areas are realized by the durations $T_1=7\pi/(3\Omega_0)$, $T_2=5\pi/(3\Omega_0)$, and $T_3=\pi/(3\Omega_0)$. The same detuning $\Delta=\delta\Omega_0$ acts during each pulse.
The second is B3r~\cite{LEVITT198661}, which compensates for amplitude errors and coincides with the SCROFULOUS sequence for an $X$ gate~\cite{Cummins_2003}, it compensates for amplitude errors, 
\begin{equation}
\mathrm{B3r}:\; \pi_{\pi/3}\; \pi_{5\pi/3}\; \pi_{\pi/3}.
\end{equation}
The third is the B3d sequence, which is robust to detuning errors,
\begin{equation}
\mathrm{B3d}:\; \pi_{2\pi/3}\; \pi_{\pi/3}\; \pi_{2\pi/3}.
\end{equation}
The performance of these sequences is compared with that of a standard $\pi$ pulse in Fig.~\ref{fig:X3}.
The CORPSE sequence broadens the high-fidelity region along the detuning axis. At $\epsilon=0$, it cancels the first-order detuning error, giving it a leading infidelity of order $\delta^4$, compared with order $\delta^2$ for a single $\pi$ pulse. It does not compensate for amplitude errors.
Near the nominal point, B3r suppresses amplitude-error sensitivity at the expense of increased detuning sensitivity, whereas B3d suppresses detuning-error sensitivity at the expense of increased amplitude sensitivity..
None of these three-pulse sequences simultaneously cancels the first-order amplitude and detuning errors in the full propagator.

\subsection{Composite sequences of five pulses}

The five-pulse symmetric sequences read
\be\label{X5sym}
\pi_{\phi_1} \pi_{\phi_2} \pi_{\phi_3} \pi_{\phi_2} \pi_{\phi_1} .
\ee
They allow us to cancel both first derivatives and the mixed second derivative,
\be
D_{1,0} \mathcal{U} = D_{0,1} \mathcal{U} = D_{1,1} \mathcal{U} = 0.
\ee
The condition $\mathcal{U}_{12} = \mathcal{U}_{21} = -i$ leads to
\be
\phi_3 = 2\phi_2 - 2\phi_1 \pmod{2\pi}.
\ee
Half of the derivatives of the propagator elements vanish due to the anagram phases.
The nonzero ones read
\begin{subequations}
\begin{align}
D_{1,0} \mathcal{U}_{11} &= -\frac{\pi}{2} \left[ 1+2\cos(\phi_1)+2\cos(2\phi_1-\phi_2) \right]  ,\\
D_{0,1} \mathcal{U}_{11} &= i \left[ 1+2\cos(\phi_1)-2\cos(2\phi_1-\phi_2) \right]  .
\end{align}
\end{subequations}
These are canceled for two sets of phases,
\begin{subequations}\label{CP5}
\begin{align}
X5a &:\ (\phi_1,\phi_2,\phi_3) = (\tfrac23,-\tfrac16,\tfrac13)\pi, \label{CP5a} \\
X5b &:\ (\phi_1,\phi_2,\phi_3) = (\tfrac23,\tfrac56,\tfrac13)\pi. \label{CP5b} 
\end{align}
\end{subequations}
Because $(\phi_1,\phi_2,\phi_3)$ and $(-\phi_1,-\phi_2,-\phi_3)$ generate the same fidelity landscape, there are two additional solutions, which are redundant and not listed. 
We also note that if $(\phi_1,\phi_2,\phi_3)$ is a solution, then $(\phi_1,\phi_2+\pi,\phi_3)$ is also a solution, although the two generate different fidelity landscapes.

These CPs are related to the universal CPs \cite{PhysRevLett.113.043001}
\begin{subequations} \label{CP5universal}
\begin{align}
\text{U5a:}&\ \pi_{0} \pi_{5\pi/6} \pi_{\pi/3} \pi_{5\pi/6} \pi_{0} , \\
\text{U5b:}&\ \pi_{0} \pi_{\pi/6} \pi_{5\pi/3} \pi_{\pi/6} \pi_{0} , 
\end{align}
\end{subequations}
by a simple phase transformation.
With the Hamiltonian and phase convention used here, the nominal values ($\epsilon=\delta=0$) of $\mathcal{U}_{12}$ for U5a and U5b are $e^{\pi i/6}$ and $e^{5\pi i/6}$, respectively. A common shift $\chi$ of all pulse phases multiplies $\mathcal{U}_{12}$ by $e^{i\chi}$. The choices $\chi=-2\pi/3$ for U5a and $\chi=2\pi/3$ for U5b therefore give the target value $-i$. The first gives the phase-reversed form of X5a in Eq.~\eqref{CP5a}, with the same target and fidelity landscape in this model; the second gives X5b in Eq.~\eqref{CP5b}, modulo $2\pi$.
The five-pulse derivative calculation thus recovers known phase families. After conversion to the present phase convention and nominal-axis alignment, U5b/X5b also gives the Knill five-pulse rotation~\cite{Ryan_2010,qile}.

For comparison, we also consider two other composite five-pulse X gates, albeit with an error suppression in a single parameter.

The BB1 sequence of Wimperis \cite{NRPBCP} is
\be\label{BB1}
\text{BB1}:\ \pi_{0}\pi_{\zeta}(2\pi)_{3\zeta}\pi_{\zeta},
\ee
where $\zeta = \arccos(-\frac14)$. 
We also test the B5 pulse of Torosov and Vitanov \cite{Torosov2011},
\be\label{B5}
\text{B5:}\ \pi_{\frac45\pi} \pi_{0} \pi_{\frac25\pi} \pi_{0} \pi_{\frac45\pi} ,
\ee
which is phase-shifted from the original pulse of Ref.~\cite{Torosov2011} by $-4\pi/5$ in order to produce the target propagator \eqref{targetX}.

\begin{figure}
\includegraphics[width=0.99\columnwidth]{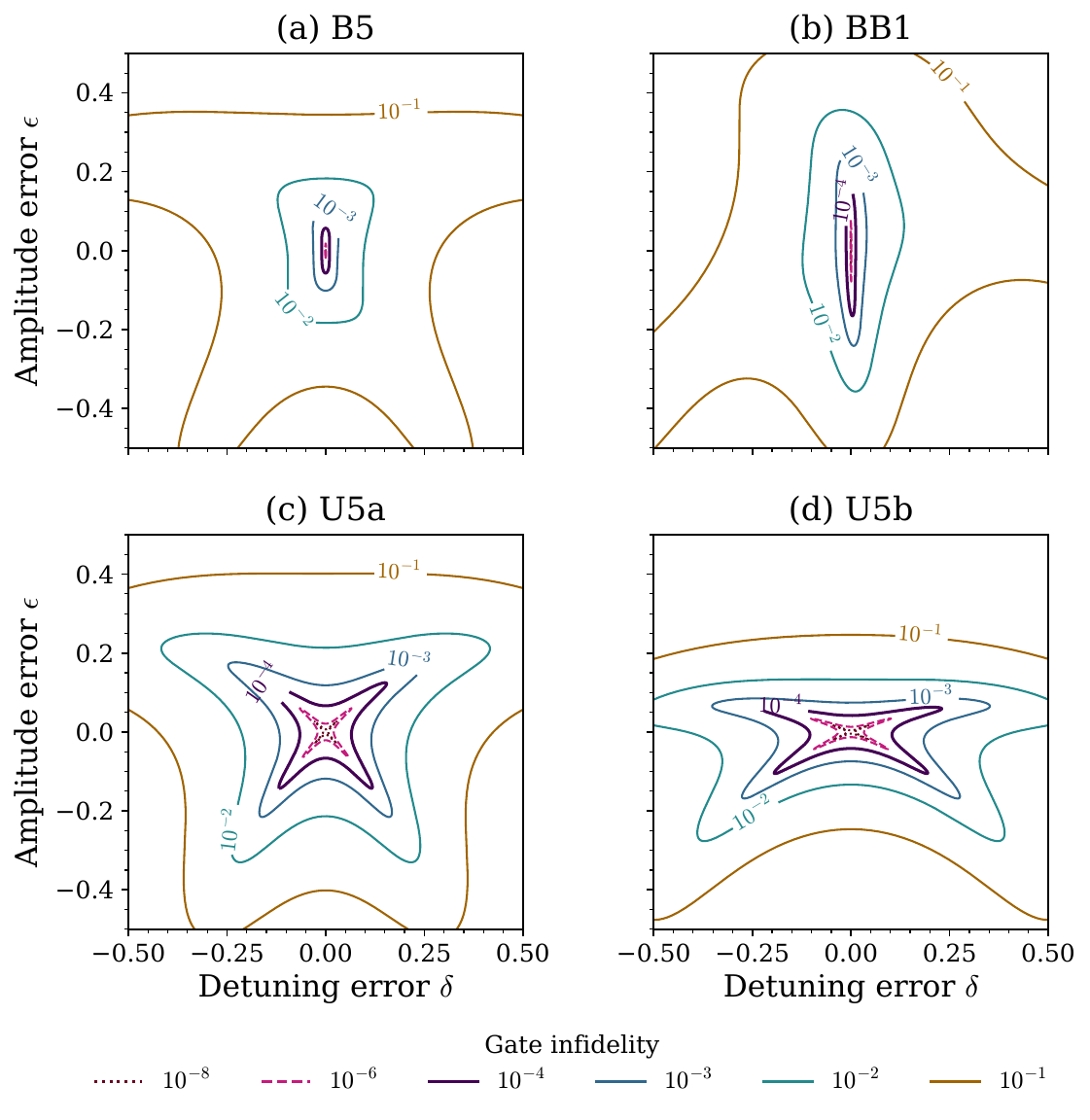} 

    \caption{(Color online)
Average gate infidelity for the $X$ gate versus the detuning error $\delta$ and the Rabi frequency error $\epsilon$ for (a) B5 pulse, (b) BB1 pulse, (c) U5a/X5a pulse, (d) U5b/X5b pulse.
Contours show infidelities $10^{-8}$, $10^{-6}$, $10^{-4}$, $10^{-3}$, $10^{-2}$, and $10^{-1}$ where resolved }
    \label{fig:X5}
\end{figure}

The fidelity profiles generated by the B5, BB1, and the universal pulses \eqref{CP5a} and \eqref{CP5b} are plotted in Fig.~\ref{fig:X5}.
The B5 and BB1 pulses compensate errors in the Rabi frequency only. 
The U5a/X5a and  U5b/X5b pulses generate broad high-fidelity ranges that are resistant to both detuning and Rabi frequency errors.
U5a/X5a outperform U5b/X5b with respect to Rabi frequency errors, whereas the opposite occurs versus detuning errors.

In addition to these error landscapes, for rectangular pulses, the propagator depends on the products $\Omega T$ and $\Delta T$. A common fractional duration error $\eta$, defined by $T\to T(1+\eta)$ at fixed Rabi frequency and detuning, can therefore be represented in the same fidelity landscape by the effective error coordinates
\begin{equation}
\epsilon_{\mathrm{eff}}=(1+\epsilon)(1+\eta)-1,
\qquad
\delta_{\mathrm{eff}}=\delta(1+\eta).
\label{eq:duration_errors}
\end{equation}
At exact resonance, $\delta=0$, a duration error moves along the amplitude-error axis, with $\epsilon_{\mathrm{eff}}=\eta$ when $\epsilon=0$. At nonzero detuning, it follows the path specified by Eq.~\eqref{eq:duration_errors}. Thus the amplitude--detuning robustness shown in Fig.~\ref{fig:X5} also implies robustness to a common duration error, provided the effective coordinates remain in the high-fidelity region.

\subsection{Longer symmetric X sequences}

The symmetric ansatz extends the full-unitary construction to $N=7,9,11,13$ pulses. We specify each sequence by its first $(N+1)/2$ phases, including the central pulse; the remaining phases follow by reflection. All constituent pulses have nominal area $\pi$. The universal comparators are shifted to match the nominal $X$ axis and are evaluated with the same gate-infidelity metric.

\subsubsection{Seven pulses}
\label{X7}
With seven pulses, cancellation of $D_{1,0}\mathcal U$, $D_{0,1}\mathcal U$, and $D_{1,1}\mathcal U$ leaves a choice between targeting $D_{2,0}\mathcal U$ and $D_{0,2}\mathcal U$. The two choices yield X7a and X7b, respectively. Defining
\begin{equation}
\xi=\arccos\!\left(\frac{3+\sqrt{61}}{16}\right),
\end{equation}
their half-phase vectors are
\begin{align}
\text{X7a}:\quad &(\pi-\xi,\,7\pi/3-2\xi,\,8\pi/3-3\xi,\notag\\
&\hspace{36mm}5\pi/3-4\xi),\label{X7a}\\
\text{X7b}:\quad &(2\pi-\xi,\,7\pi/3-2\xi,\,5\pi/3-3\xi,\notag\\
&\hspace{36mm}5\pi/3-4\xi).\label{X7b}
\end{align}

These solutions belong to the symmetric seven-pulse family
introduced by Jones~\cite{Jones_2013_NOT}. Denoting the full
phase vector in Eq.~(58) of that reference by
$\boldsymbol{\phi}_J(\alpha)$, the relation in the present
phase convention is
\begin{align*}
\boldsymbol{\phi}_{X7a}
&=\pi\mathbf{1}-\boldsymbol{\phi}_J(\xi),\\
\boldsymbol{\phi}_{X7b}
&=\pi\mathbf{1}-\boldsymbol{\phi}_J(\xi-\pi),
\end{align*}
with all phases understood modulo $2\pi$. Here $\mathbf{1}$
is the seven-component vector of ones. The derivative
conditions therefore select particular members of this
known family. Their parameter values differ from those of
the Jones 7b and 7c representatives included in
Table~\ref{table:resource_comparison}.

\begin{figure}
\includegraphics[width=0.99\columnwidth]{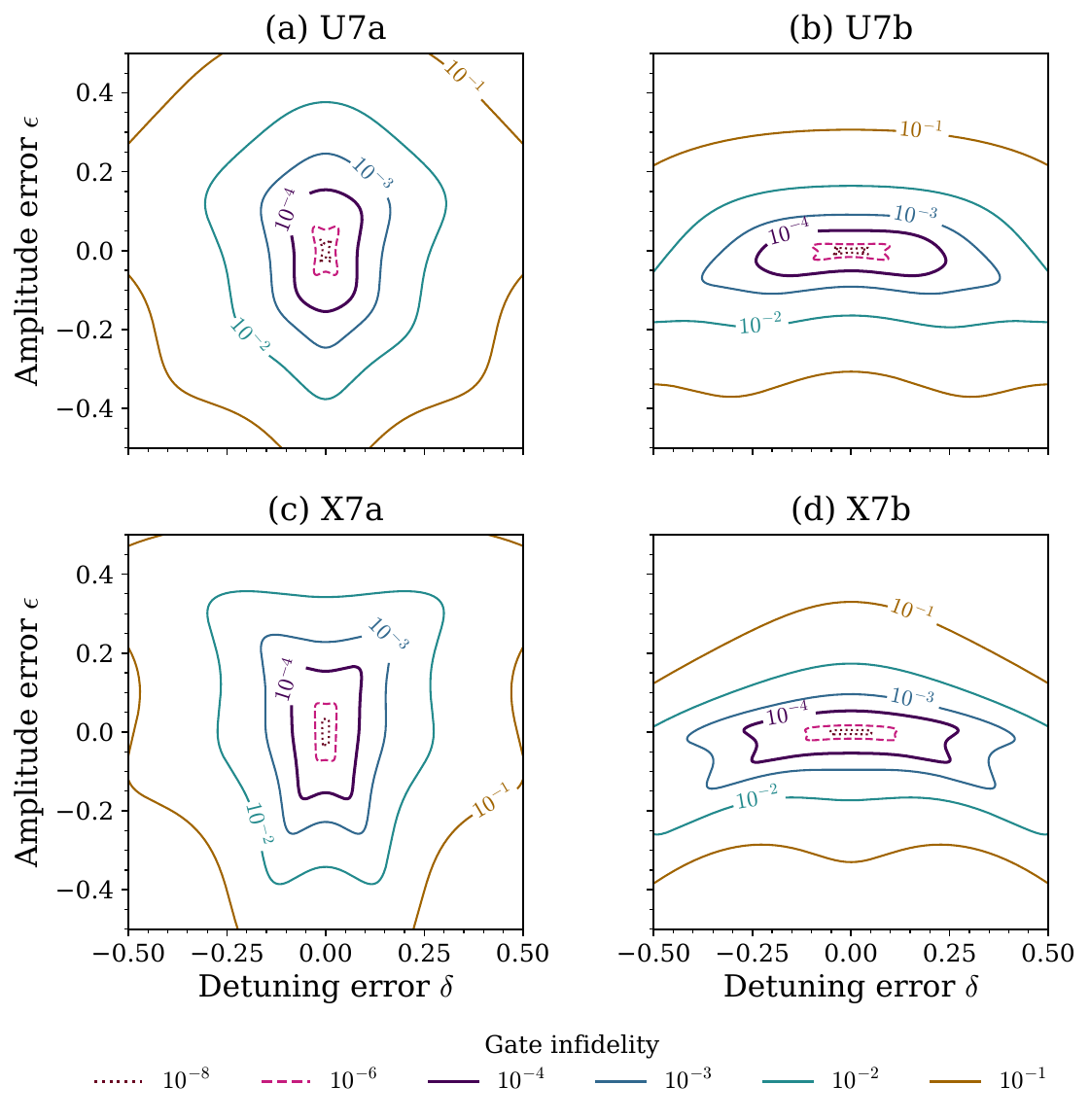}
\caption{Gate infidelity versus detuning error $\delta$ and amplitude error $\epsilon$ for seven-pulse sequences: (a) U7a, (b) U7b, (c) X7a, and (d) X7b. The contours show $10^{-8}$ (dotted dark red), $10^{-6}$ (dashed magenta), $10^{-4}$ (emphasized purple), $10^{-3}$ (blue), $10^{-2}$ (teal), and $10^{-1}$ (ochre), where resolved. All four controls have area $7\pi$ and duration $7\pi/\Omega_0$.}
\label{fig:X7best}
\end{figure}

The corresponding universal comparator vectors are
\begin{align}
\text{U7a}:\quad &(5/12,\,1/2,\,19/12,\,0)\pi,\\
\text{U7b}:\quad &(7/12,\,3/2,\,17/12,\,0)\pi.
\end{align}
Figure~\ref{fig:X7best} compares the universal and derivative-designed seven-pulse landscapes. The two derivative choices select different error sensitivities: X7a favors amplitude tolerance, whereas X7b favors detuning tolerance. Thus even at fixed pulse count and area, the derivative conditions provide a choice of robustness profile.

Table~\ref{table:resource_comparison} also includes the seven-pulse Jones 7a, 7b, and 7c constructions~\cite{Jones_2013_NOT}. All of these seven-pulse controls use area $7\pi$ and duration $7\pi/\Omega_0$. On $|\epsilon|,|\delta|\leq0.15$, the mean infidelity of X7a, $2.62\times10^{-4}$, is close to those of U7a and Jones 7b, $2.52\times10^{-4}$ and $2.66\times10^{-4}$, and below that of Jones 7a, $5.63\times10^{-4}$. X7b and U7b have means $1.35\times10^{-3}$ and $1.36\times10^{-3}$, respectively, while Jones 7c gives $1.18\times10^{-3}$. The seven-pulse derivative solutions therefore offer different robustness profiles rather than a uniform reduction of the mean error relative to every analytical reference.

\subsubsection{Nine pulses}
\label{X9}
\begin{figure}
\includegraphics[width=0.99\columnwidth]{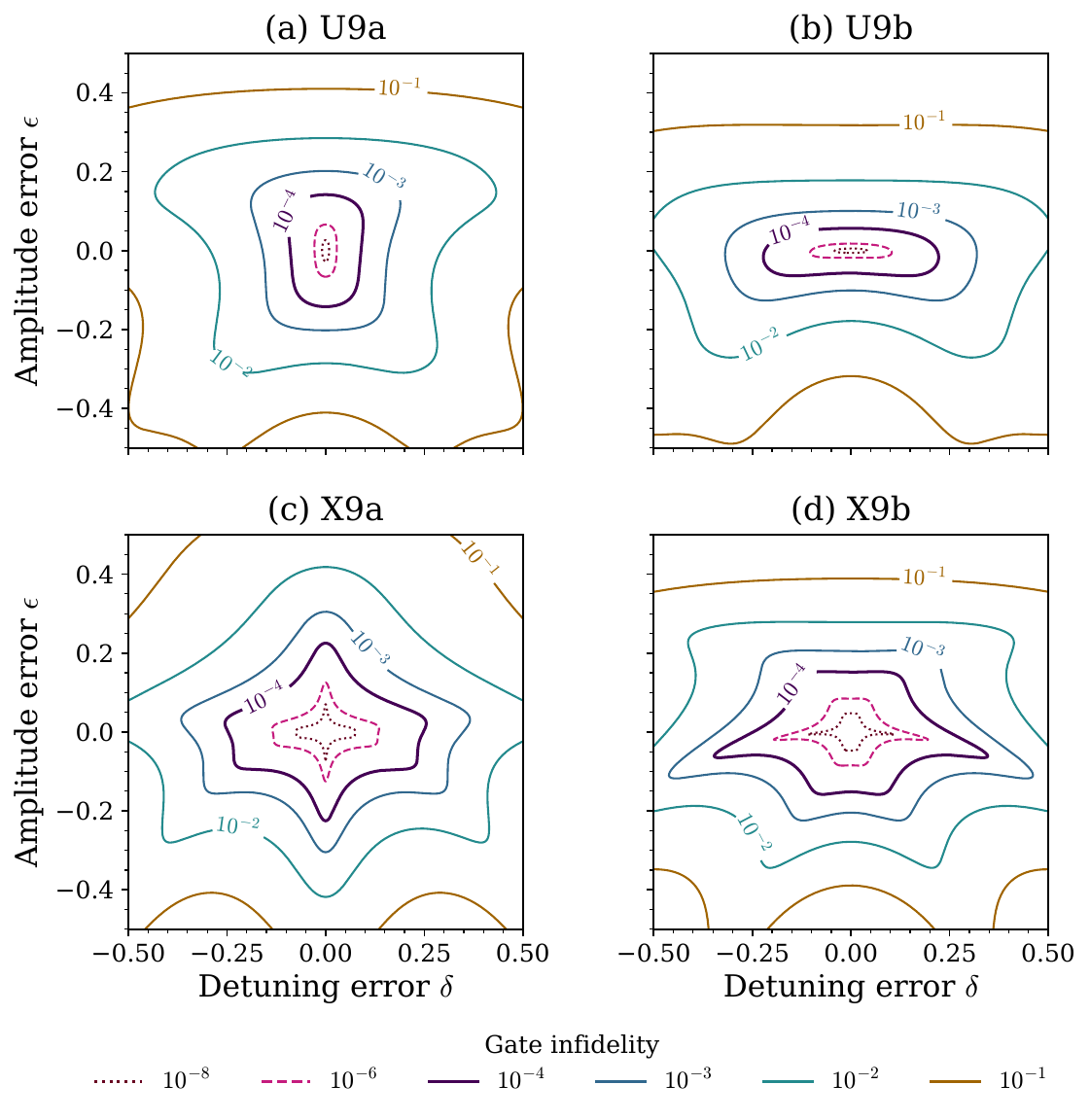}
\caption{Gate infidelity versus detuning error $\delta$ and amplitude error $\epsilon$ for nine-pulse sequences: (a) U9a, (b) U9b, (c) X9a, and (d) X9b. Contour levels, colors, and line styles follow Fig.~\ref{fig:X7best}.}
\label{fig:X9best}
\end{figure}

For nine pulses we target all first- and second-order derivatives,
\begin{equation}
D_{1,0}\mathcal U=D_{0,1}\mathcal U=D_{2,0}\mathcal U
=D_{1,1}\mathcal U=D_{0,2}\mathcal U=0.
\end{equation}
Two solutions have half-phase vectors
\begin{align}
\text{X9a}:\quad &(\xi_1+\pi,\,\xi_2,\,2\xi_1,\,5\xi_1-\xi_2,\,4\xi_1),\\
\text{X9b}:\quad &(\xi_1+\pi,\,\xi_2+\pi,\,2\xi_1,\notag\\
&\hspace{22mm}5\xi_1-\xi_2-\pi,\,4\xi_1),
\end{align}
where $\xi_1=\arctan\sqrt{15}$ and $\xi_2=\arctan(\sqrt{15}/9)$. The universal comparators are
\begin{align}
\text{U9a}:\quad &(4/3,\,35/24,\,3/4,\,35/24,\,5/3)\pi,\\
\text{U9b}:\quad &(2/3,\,37/24,\,5/4,\,37/24,\,1/3)\pi.
\end{align}
Figure~\ref{fig:X9best} compares X9a and X9b with the aligned universal sequences U9a and U9b. The derivative-designed sequences give lower mean infidelities and broader $10^{-4}$ regions in the central square $|\epsilon|,|\delta|\leq0.15$; Table~\ref{table:resource_comparison} quantifies these differences.

The nine-pulse solutions also allow a matched-resource comparison with the Jones S9 construction~\cite{Jones_2013_NOT}: all three use area $9\pi$ and duration $9\pi/\Omega_0$. On the same square, X9a and X9b have mean infidelities $5.52\times10^{-5}$ and $2.80\times10^{-5}$, compared with $1.43\times10^{-4}$ for Jones S9. As reported in Table~\ref{table:resource_comparison}, the factors of approximately 2.6 and 5.1 quantify the finite-domain benefit of our solutions at the same resources and local total cancellation order.

\subsubsection{Eleven and thirteen pulses}
\label{X11}
\begin{figure}
\includegraphics[width=0.99\columnwidth]{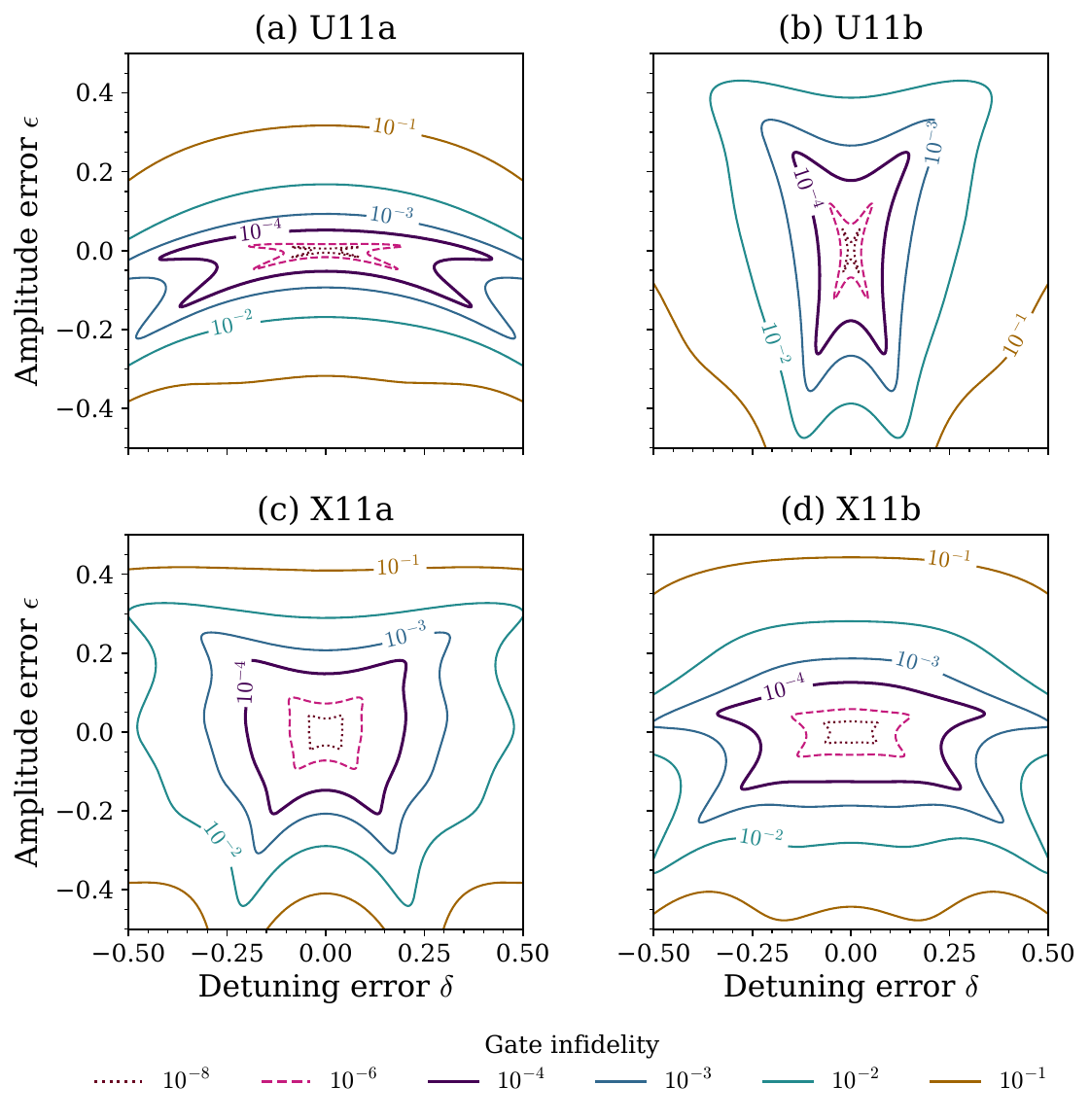}
\caption{Gate infidelity versus detuning error $\delta$ and amplitude error $\epsilon$ for eleven-pulse symmetric sequences: (a) U11a, (b) U11b, (c) X11a, and (d) X11b. The contour levels, colors, and line styles follow Fig.~\ref{fig:X7best}.}
\label{fig:X11best}
\end{figure}

At eleven pulses, the same first- and second-order conditions are supplemented by $D_{2,1}\mathcal U=0$ for X11a or $D_{1,2}\mathcal U=0$ for X11b. The numerical half-phase vectors, rounded to five decimals in units of $\pi$, are
\begin{align}
\text{X11a}:\quad &(0.55329,\,0.80089,\,0.70913,\notag\\&\hspace{8mm}1.44638,\,0.68093,\,0.39214),\\
\text{X11b}:\quad &(1.55329,\,0.80089,\,1.70913,\notag\\&\hspace{8mm}1.44638,\,1.68093,\,0.39214).
\end{align}
For comparison we use
\begin{align}
\text{U11a}:\quad &(5/12,\,4/3,\,5/4,\,1/3,\,1/2,\,0)\pi,\\
\text{U11b}:\quad &(5/12,\,1/3,\,5/4,\,4/3,\,1/2,\,1)\pi.
\end{align}
Figure~\ref{fig:X11best} shows the resulting full-gate landscapes. Since all four sequences have the same count, area and duration, their differences isolate phase selection within the stated model. 

For thirteen pulses, both mixed conditions $D_{2,1}\mathcal U=0$ and $D_{1,2}\mathcal U=0$ are added to the first- and second-order conditions. Two numerical half-phase vectors are
\begin{align}
\text{X13a}:\quad &(0.53248,\,0.50726,\,1.29153,\notag\\&\hspace{8mm}0.44431,\,0.73020,\,0.48083,\,1.75637)\pi,\label{X13}\\
\text{X13b}:\quad &(0.53248,\,1.50726,\,1.29153,\notag\\&\hspace{8mm}1.44431,\,0.73020,\,1.48083,\,1.75637)\pi.
\end{align}
The corresponding universal comparator vectors are
\begin{align}
\text{U13a}:\quad &(1/2,\,7/8,\,9/4,\,23/24,\,5/6,\notag\\
&\hspace{32mm}49/24,\,7/12)\pi,\\
\text{U13b}:\quad &(1/2,\,15/8,\,9/4,\,47/24,\,5/6,\notag\\
&\hspace{32mm}25/24,\,7/12)\pi.
\end{align}

The additional mixed conditions shape the joint-error response but leave the pure third derivatives untargeted, i.e. they do not establish cancellation through total order three. The performance of the pulses can be seen in Table~\ref{table:resource_comparison}. We also note that for $X11$ and $X13$, the numerical phase coefficients satisfy the targeted derivative conditions approximately, as they are not derived analytically.

We conclude the presentation of the symmetric sequences in Figs.~\ref{fig:X7best}, \ref{fig:X9best} and \ref{fig:X11best} obtained through the cancellation of derivatives of propagator elements by noting a clear trend: as the length of the pulses (and thus the total duration) increases, the robustness window broadens. This is an expected trade-off inherent to CP, that is directly explored in Fig~\ref{fig:normalized_resources}.
This is also valid for the  universal CP which we adapted to produce the desired X gates.
Clearly though they do not provide as broad error domains as the newly derived X$n$ sequences.

\subsection{Numerical variable-area X gates}
\label{Sec:variable_X}

Variable-area designs use phases $\phi_k$ and amplitudes $a_k=\Omega_k/\Omega_0$ at equal segment duration, $\Omega_0T=\pi$. The numerical $X$ and Hadamard catalogues contain sequences with $N=5,7,9,11,13,15$, with paired central-accuracy (c) and width-oriented (w) designs at $N=13,15$. Tables~\ref{table: paramsX} and \ref{table: params} list the pulse areas and phases of the numerical controls, rounded to five decimal places. Full-precision controls and benchmark data are available in the reproduction repository~\cite{my_github_repo}.

For comparisons between sequences with different amplitudes, we fix the same physical peak nominal Rabi frequency $\Omega_{\max}$ and preserve every pulse's rotation angle by rescaling amplitudes and inversely rescaling durations. Let $p=\max_k a_k$ denote the peak amplitude in units of the construction reference $\Omega_0$. The detuning coordinate for these comparisons is
\begin{equation}
\delta_c=\frac{\Delta}{\Omega_{\max}},\qquad
\Omega_{\max}=p\Omega_0,\qquad
\delta=p\delta_c.
\label{eq:common_peak_detuning}
\end{equation}
When different sequences are brought to the same physical peak, their reference frequencies $\Omega_0=\Omega_{\max}/p$ can differ. Equal values of the native coordinate $\delta=\Delta/\Omega_0$ would then correspond to different physical detuning offsets. Using a shared $\delta_c$ instead compares the gates at the same $\Delta$ and the same peak drive. For the constant-amplitude sequences, $p=1$ and $\delta_c=\delta$.

The $X_{nc}$ sequences are derived by minimizing either Eq.~\eqref{eq:grid_infidelity} or Eq.~\eqref{eq:width_loss}. All $X_{nc}$ CP except X5c exhibit, up to numerical precision, no error at the nominal point, with X5c having nominal infidelity of $8.27\times10^{-5}$. The wider pair X13w and X15w use Eq.\eqref{eq:width_loss} on the common-peak square of half-width 0.5, and are bound by the same optimization constraints $0.05\leq a_k\leq2$, and $\sum_k a_k\leq N$ as their $Xnc$ counterparts. All optimization and refinement procedures can be found in the reproduction materials~\cite{my_github_repo}.

\begin{figure*}[tp]
\includegraphics[width=0.99\textwidth]{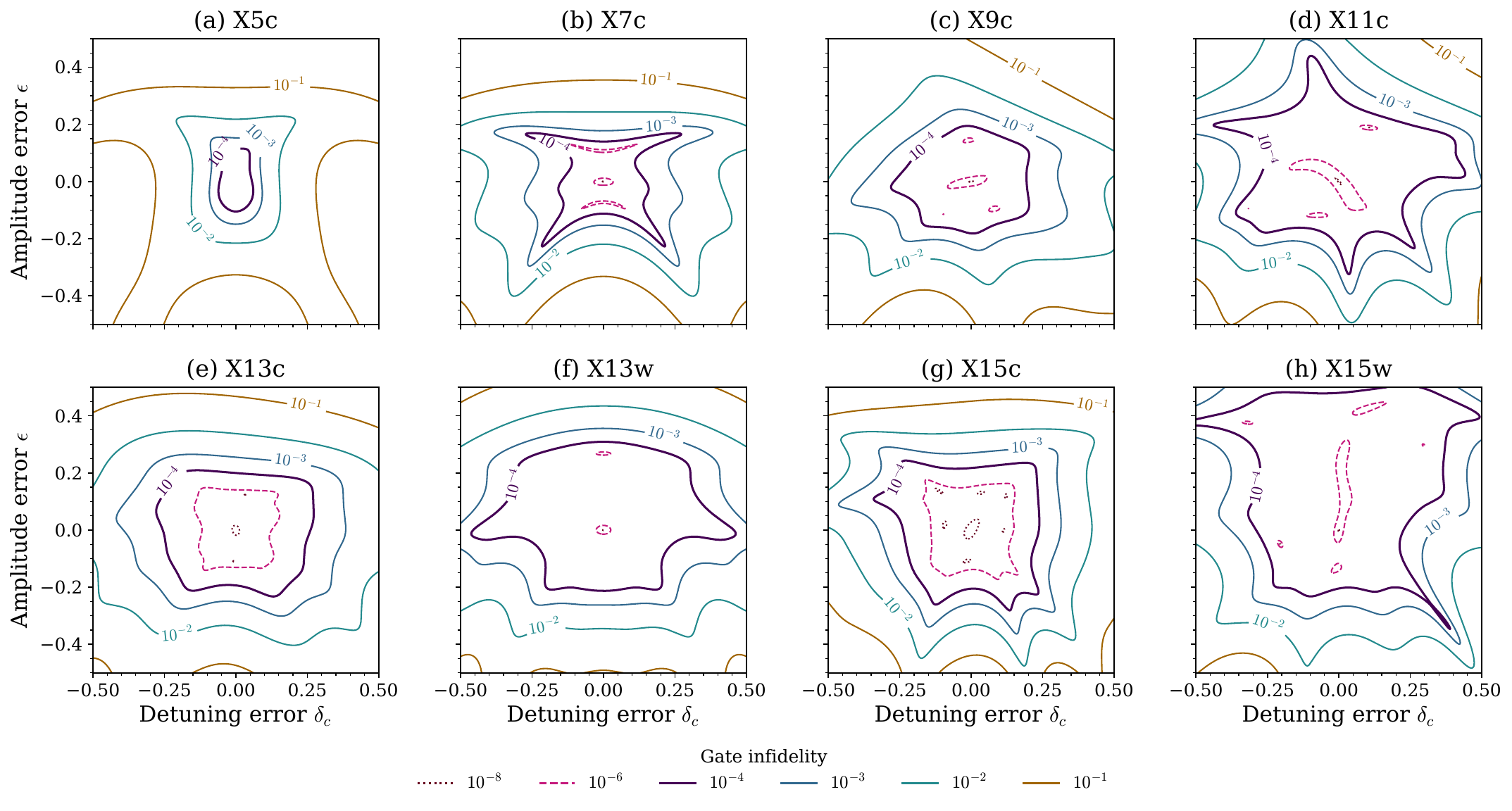}
\caption{Variable-area $X$ landscapes in common-peak coordinates: (a--d) X5c, X7c, X9c, X11c; (e,f) X13c, X13w; (g,h) X15c, X15w. The contour levels are $10^{-8}$, $10^{-6}$, $10^{-4}$, $10^{-3}$, $10^{-2}$, and $10^{-1}$ where resolved. The emphasized $10^{-4}$ contour measures threshold width, while the inner contours resolve central error suppression. The same level colors and line styles are used throughout the landscape figures.}
\label{fig:Xnonsym}
\end{figure*}

Figure~\ref{fig:Xnonsym} illustrates the distinction between
central accuracy and threshold coverage. $X13c$ and $X15c$ lower the mean infidelity on $B_{0.15}$ to $7.74\times10^{-7}$ and $2.57\times10^{-7}$ (with $B_r=\{(\epsilon,\delta_c):|\epsilon|,|\delta_c|\leq r\}$). Whereas their width-oriented counterparts increase the connected \(10^{-4}\) fractions of $B_{0.5}$ from \(19.64\%\) to \(29.45\%\) and from $21.30\%$ to $43.57\%$, respectively, at the expense of higher central mean infidelities.

The central mean infidelity decreases along the numerical
$X_{nc}$ catalogue as the pulse count increases. At
$N=7,9,11,13$, $X_{nc}$ designs have lower means
than the symmetric derivative-designed controls at the
same pulse count
(Table~\ref{table:resource_comparison}). X5c, however, has a higher mean infidelity than X5a despite its larger area. To perform a proper comparison, we define the normalized pulse area.
\begin{equation}
\Phi=\sum_k\Omega_kT_k,\qquad
\mathcal R_A=\frac{\Phi}{\theta},
\label{eq:resource_costs}
\end{equation}
where $\theta=\pi$ for $X$ and $\pi/2$ for the Hadamard primitive. A single resonant target rotation thus has unit cost. Normalized pulse area accounts for sequences that use few segments but require large rotation angles. For example the reduced CORPSE-in-BB1 uses six segments but has an area of $8.333\pi$ for the $X$ target.

Table~\ref{table:resource_comparison} presents all the constant-amplitude, universal, derivative-designed, numerical and higher-order analytical controls. The mean, sampled maximum and threshold fraction on $B_{0.15}$ characterize the calibration window while connected fraction on $B_{0.5}$ characterizes larger-error tolerance. The sampled maximum reports the largest infidelity among the evaluated points. Fig.~\ref{fig:normalized_resources}(a) compares the central mean infidelities against normalized pulse area.
This trend is not monotonic though, $X11b$ has a higher mean than
$X9b$, and $X13a$ has a slightly higher mean than $X11a$. Most clearly this trend is visible for the $Xnc$ sequences.

\begin{table*}[tp]
\centering\small
\setlength{\tabcolsep}{4pt}
\begin{ruledtabular}
\begin{tabular}{lrrrrrr}
Sequence & $N$ & $\mathcal R_A$ & $\overline{\mathcal I}_{0.15}$ & $\mathcal I_{\max,0.15}$ & $f_4(0.15)$ (\%) & $f_4^{(0)}(0.5)$ (\%) \\
Single $X$ & 1 & 1.000 & $1.7\times10^{-2}$ & $5.2\times10^{-2}$ & 0.33 & 0.03 \\
SCROFULOUS & 3 & 3.000 & $2.0\times10^{-2}$ & $7.5\times10^{-2}$ & 1.81 & 0.16 \\
SCORBUTUS & 5 & 5.000 & $1.9\times10^{-3}$ &
$1.5\times10^{-2}$ & 18.87 & 1.70 \\
CORPSE & 3 & 4.333 & $1.4\times10^{-2}$ & $4.8\times10^{-2}$ & 4.67 & 0.44 \\
B3d & 3 & 3.000 & $4.7\times10^{-2}$ & $1.5\times10^{-1}$ & 1.81 & 0.17 \\
B5 & 5 & 5.000 & $7.7\times10^{-3}$ & $2.7\times10^{-2}$ & 2.21 & 0.20 \\
BB1 & 4 & 5.000 & $6.0\times10^{-3}$ & $3.8\times10^{-2}$ & 7.23 & 0.67 \\
rCinSK & 5 & 8.333 & $2.7\times10^{-3}$ & $4.0\times10^{-2}$ & 17.73 & 1.59 \\
rCinBB & 6 & 8.333 & $1.0\times10^{-3}$ & $1.6\times10^{-2}$ & 30.77 & 2.83 \\
rSKinsC & 6 & 6.333 & $3.8\times10^{-3}$ & $5.4\times10^{-2}$ & 18.88 & 1.70 \\
Jones 7a & 7 & 7.000 & $5.6\times10^{-4}$ & $7.2\times10^{-3}$ & 42.23 & 5.03 \\
Jones 7b & 7 & 7.000 & $2.7\times10^{-4}$ & $2.4\times10^{-3}$ & 48.60 & 4.61 \\
Jones 7c & 7 & 7.000 & $1.2\times10^{-3}$ & $7.2\times10^{-3}$ & 35.01 & 4.57 \\
U5a / X5a & 5 & 5.000 & $5.1\times10^{-4}$ & $2.9\times10^{-3}$ & 34.95 & 3.15 \\
U5b / X5b & 5 & 5.000 & $2.7\times10^{-3}$ & $1.6\times10^{-2}$ & 29.81 & 3.14 \\
U7a & 7 & 7.000 & $2.5\times10^{-4}$ & $2.1\times10^{-3}$ & 50.93 & 4.61 \\
U7b & 7 & 7.000 & $1.4\times10^{-3}$ & $8.3\times10^{-3}$ & 35.25 & 4.57 \\
U9a & 9 & 9.000 & $1.8\times10^{-4}$ & $1.1\times10^{-3}$ & 51.09 & 4.60 \\
U9b & 9 & 9.000 & $9.2\times10^{-4}$ & $5.3\times10^{-3}$ & 37.70 & 4.57 \\
U11a & 11 & 11.000 & $1.3\times10^{-3}$ & $8.5\times10^{-3}$ & 35.68 & 7.75 \\
U11b & 11 & 11.000 & $2.1\times10^{-4}$ & $1.9\times10^{-3}$ & 58.38 & 7.91 \\
U13a & 13 & 13.000 & $2.2\times10^{-4}$ & $1.6\times10^{-3}$ & 56.62 & 6.69 \\
U13b & 13 & 13.000 & $1.5\times10^{-3}$ & $9.2\times10^{-3}$ & 35.91 & 6.59 \\
X7a & 7 & 7.000 & $2.6\times10^{-4}$ & $2.9\times10^{-3}$ & 54.44 & 5.23 \\
X7b & 7 & 7.000 & $1.3\times10^{-3}$ & $9.9\times10^{-3}$ & 34.92 & 5.19 \\
X9a & 9 & 9.000 & $5.5\times10^{-5}$ & $7.7\times10^{-4}$ & 82.91 & 11.18 \\
X9b & 9 & 9.000 & $2.8\times10^{-5}$ & $4.5\times10^{-4}$ & 93.25 & 11.16 \\
X11a & 11 & 11.000 & $1.2\times10^{-5}$ & $1.1\times10^{-4}$ & 99.75 & 12.55 \\
X11b & 11 & 11.000 & $4.7\times10^{-5}$ & $4.9\times10^{-4}$ & 82.06 & 12.48 \\
X13a & 13 & 13.000 & $1.2\times10^{-5}$ & $1.2\times10^{-4}$ & 99.90 & 17.63 \\
X13b & 13 & 13.000 & $3.3\times10^{-6}$ & $4.1\times10^{-5}$ & 100.00 & 17.72 \\
Jones S9 & 9 & 9.000 & $1.4\times10^{-4}$ & $2.0\times10^{-3}$ & 70.33 & 8.99 \\
ASBO-9$+$ & 9 & 9.000 & $1.0\times10^{-3}$ & $2.3\times10^{-2}$ & 52.20 & 7.75 \\
ASBO-9$-$ & 9 & 9.000 & $5.4\times10^{-4}$ & $1.8\times10^{-2}$ & 66.16 & 7.84 \\
$\mathrm{CinAP3}_{1}$ & 9 & 16.333 & $8.2\times10^{-3}$ & $7.6\times10^{-2}$ & 13.61 & 1.26 \\
$\mathrm{CinPD4}_{2}$ & 11 & 20.333 & $9.9\times10^{-3}$ & $5.8\times10^{-2}$ & 11.26 & 1.06 \\
$\mathrm{CinAP5}_{4}$ & 13 & 24.333 & $1.5\times10^{-2}$ & $1.2\times10^{-1}$ & 9.35 & 0.88 \\
$\mathrm{CinPD6}_{3}$ & 15 & 28.333 & $1.4\times10^{-2}$ & $7.9\times10^{-2}$ & 8.67 & 1.09 \\
X5c & 5 & 6.992 & $1.9\times10^{-3}$ & $1.1\times10^{-2}$ & 24.48 & 2.20 \\
X7c & 7 & 6.926 & $5.6\times10^{-5}$ & $8.9\times10^{-4}$ & 84.63 & 8.92 \\
X9c & 9 & 8.945 & $1.2\times10^{-5}$ & $2.6\times10^{-4}$ & 98.69 & 12.91 \\
X11c & 11 & 10.529 & $7.1\times10^{-6}$ & $4.2\times10^{-5}$ & 100.00 & 28.49 \\
X13c & 13 & 13.000 & $7.7\times10^{-7}$ & $1.2\times10^{-5}$ & 100.00 & 19.64 \\
X13w & 13 & 12.250 & $2.2\times10^{-5}$ & $5.2\times10^{-5}$ & 100.00 & 29.45 \\
X15c & 15 & 15.000 & $2.6\times10^{-7}$ & $9.5\times10^{-6}$ & 100.00 & 21.30 \\
X15w & 15 & 13.016 & $1.6\times10^{-5}$ & $5.9\times10^{-5}$ & 100.00 & 43.57 \\
\end{tabular}
\end{ruledtabular}
\caption{$X$-gate performance at common peak nominal Rabi frequency. $\mathcal R_A=\Phi/\theta$ uses $\theta=\pi$, and $N$ counts the constituent rotations in a given sequence. Means, sampled maxima and $f_4(0.15)$ refer to $B_{0.15}$; $f_4^{(0)}(0.5)$ is the fraction of $B_{0.5}$ occupied by a connected region with $\mathcal I\leq10^{-4}$. The two fractions use different, explicitly specified domains. Means and fractions use $801\times801$ midpoint grids, with boundaries additionally checked for maxima. All entries use the full-precision controls. The AP/PD entries are the published branches with the lowest independently evaluated mean at each amplitude order.}
\label{table:resource_comparison}
\end{table*}

\section{Hadamard gates}
\label{Sec:Hadamard}

For $R_\alpha(\theta)=e^{-i\theta\sigma_\alpha/2}$, the Hadamard gate follows from
\begin{equation}
H=e^{i\pi/2}R_z(\pi/2)R_x(\pi/2)R_z(\pi/2).
\end{equation}
where we assume ideal virtual $Z$ rotations~\cite{McKay_2017}, that do not alter the infidelity of the implemented $R_x(\pi/2)$. Pulse-phase errors and finite digital phase resolution are examined in Appendix~\ref{app:phase_precision}.

The Hadamard catalogue mirrors the $X$ catalogue in pulse count and c/w structure. Its numerical controls use the objectives and constant/finite-edge model specified in Sec.~\ref{Sec:derivation}. The landscapes below show rectangular-pulse performance. Appendix~\ref{app:pulse_shape} examines finite pulse edges and time-dependent detuning. We characterize the numerical designs by finite-domain accuracy without assigning them a perturbative cancellation order, as was the case for $Xna$ and $Xnb$.

\begin{figure}[tbp]
\includegraphics[width=\columnwidth]{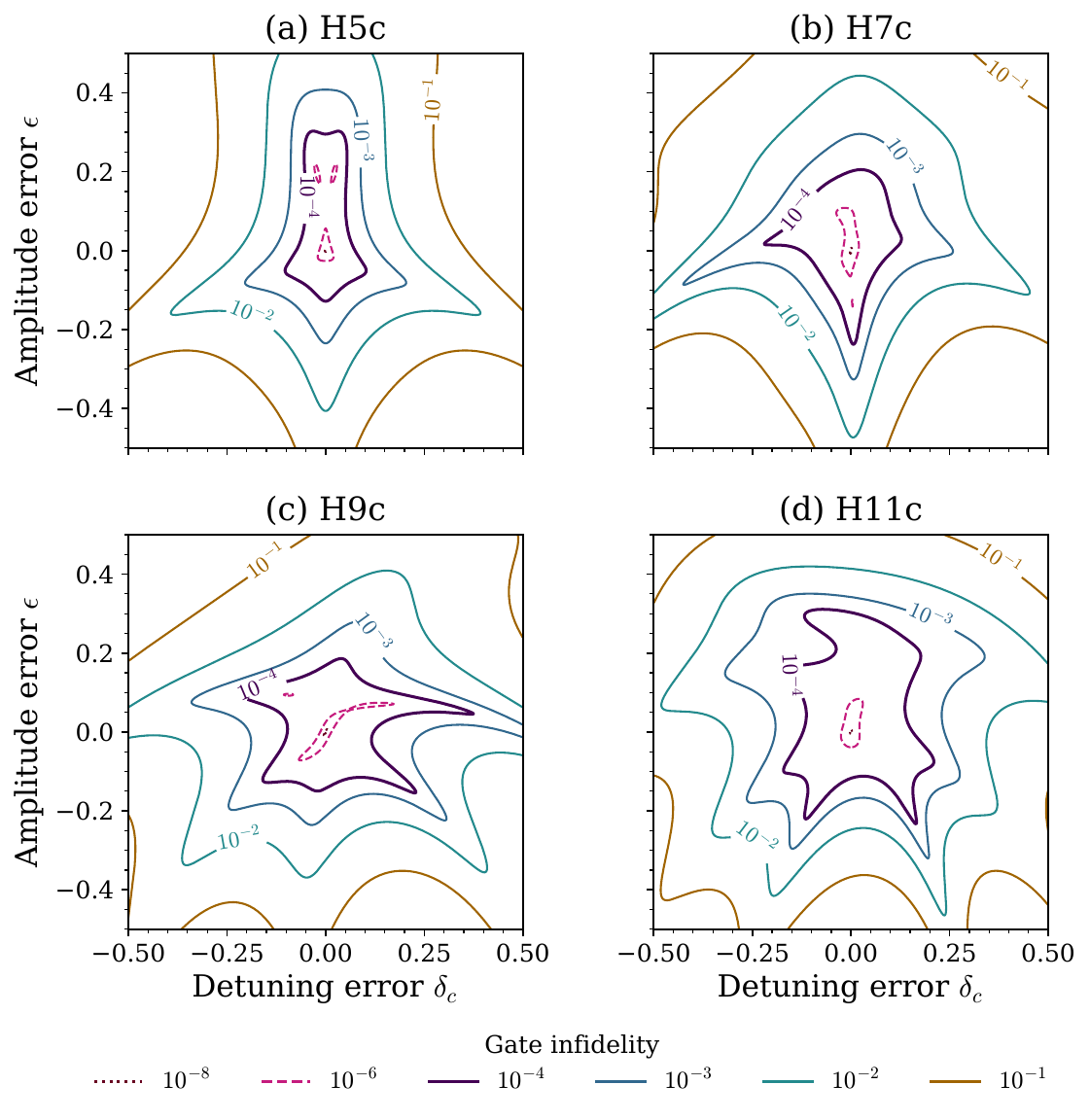}
\caption{Rectangular-pulse Hadamard-primitive landscapes for (a) H5c, (b) H7c, (c) H9c and (d) H11c. All four controls are evaluated relative to $R_x(\pi/2)$. Ideal virtual $Z$ updates complete the Hadamard gate.}
\label{fig:H37}
\end{figure}

\begin{figure}[tbp]
\includegraphics[width=\columnwidth]{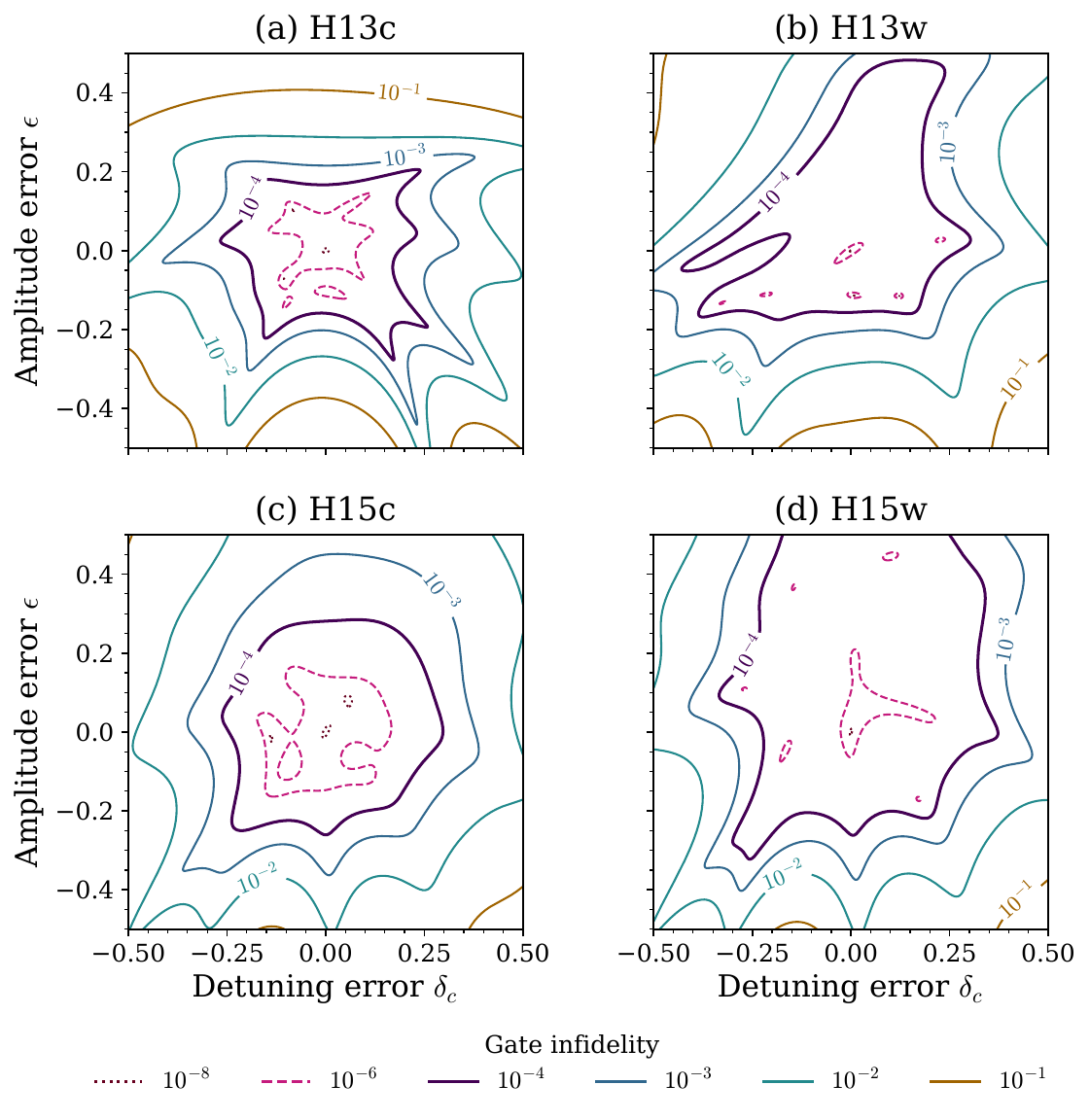}
\caption{Precision and coverage choices for the Hadamard primitive: (a,b) H13c and H13w; (c,d) H15c and H15w. Each coverage design has the same pulse count as its precision counterpart and does not exceed its area or peak bound.}
\label{fig:H715}
\end{figure}

Figures~\ref{fig:H37} and \ref{fig:H715} show the progression from the shorter Hadamard constructions to the longer central-accuracy and width-oriented designs. Among the shorter sequences, the mean infidelity on $B_{0.15}$ decreases from $6.9\times10^{-4}$ for H5c to $4.5\times10^{-5}$ for H11c. H9c provides a useful comparison with reduced CORPSE-in-BB1: its mean infidelity is approximately six times lower while its normalized pulse area is about $26\%$ smaller, although it uses nine rather than six segments (Table~\ref{table:H_comparison}). 

\begin{figure*}[tp]
\includegraphics[width=0.99\textwidth]{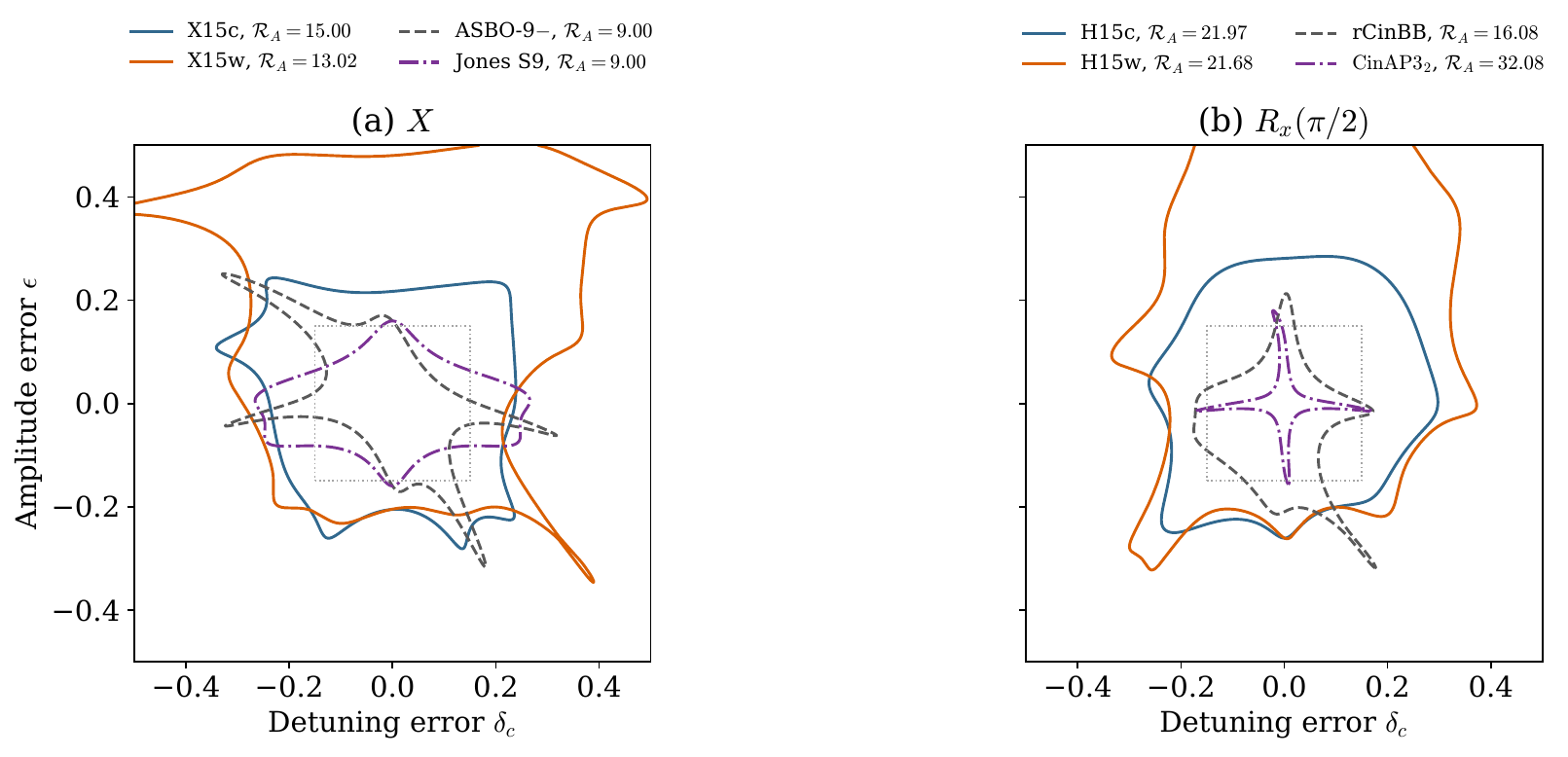}
\caption{Connected accuracy-window comparison at infidelity $10^{-4}$ for (a) $X$ and (b) the Hadamard primitive. The c/w pairs obey paired pulse-count and area bounds. The pulse area resource costs are shown in the legends. The dotted square is $B_{0.15}$. Panel (a) includes Jones S9 and ASBO-9$-$; panel (b) includes rCinBB and $\mathrm{CinAP3}_2$.}
\label{fig:higher_objectives}
\end{figure*}

The longer H13c and H15c sequences further reduce the rectangular-pulse means to $4.7\times10^{-6}$ and $8.9\times10^{-7}$. Their width-oriented counterparts redistribute this accuracy over a larger error domain, as seen in Fig.~\ref{fig:H715}: the connected $10^{-4}$ fractions of $B_{0.5}$ increase from $14.46\%$ to $23.84\%$ for the 13-pulse pair and from $22.85\%$ to $40.05\%$ for the 15-pulse pair. These wider regions come with shallower central error suppression. The paired constructions therefore offer a choice between greater accuracy within the central window and tolerance of larger amplitude and detuning offsets. Table~\ref{table:H_comparison} quantifies this trade-off and Fig.~\ref{fig:higher_objectives} makes it explicit.

\begin{table*}[tp]
\centering\small
\setlength{\tabcolsep}{4pt}
\begin{ruledtabular}
\begin{tabular}{lrrrrrr}
Sequence & $N$ & $\mathcal R_A$ & $\overline{\mathcal I}_{0.15}$ & $\mathcal I_{\max,0.15}$ & $f_4(0.15)$ (\%) & $f_4^{(0)}(0.5)$ (\%) \\
Single $R_x(\pi/2)$ & 1 & 1.000 & $5.6\times10^{-3}$ & $1.7\times10^{-2}$ & 0.94 & 0.08 \\
SCORBUTUS & 5 & 8.705 & $1.6\times10^{-3}$ &
$1.5\times10^{-2}$ & 24.17 & 2.18 \\
rCinSK & 5 & 16.080 & $4.1\times10^{-3}$ & $5.1\times10^{-2}$ & 17.32 & 1.59 \\
rCinBB & 6 & 16.080 & $4.1\times10^{-4}$ & $1.1\times10^{-2}$ & 59.04 & 6.65 \\
rSKinsC & 6 & 12.080 & $5.9\times10^{-3}$ & $7.9\times10^{-2}$ & 21.68 & 2.03 \\
$\mathrm{CinAP3}_{2}$ & 9 & 32.080 & $1.1\times10^{-2}$ & $8.8\times10^{-2}$ & 11.84 & 1.11 \\
$\mathrm{CinAP4}_{2}$ & 11 & 40.080 & $1.4\times10^{-2}$ & $9.7\times10^{-2}$ & 9.85 & 0.90 \\
$\mathrm{CinAP5}_{4}$ & 13 & 48.080 & $1.4\times10^{-2}$ & $9.0\times10^{-2}$ & 9.17 & 0.88 \\
$\mathrm{CinAP6}_{4}$ & 15 & 56.080 & $1.5\times10^{-2}$ & $9.2\times10^{-2}$ & 8.53 & 0.86 \\
H5c & 5 & 13.938 & $6.9\times10^{-4}$ & $7.7\times10^{-3}$ & 37.76 & 4.97 \\
H7c & 7 & 14.980 & $3.6\times10^{-4}$ & $1.0\times10^{-2}$ & 60.90 & 6.61 \\
H9c & 9 & 11.859 & $6.6\times10^{-5}$ & $7.8\times10^{-4}$ & 78.84 & 8.53 \\
H11c & 11 & 19.142 & $4.5\times10^{-5}$ & $5.4\times10^{-4}$ & 87.83 & 12.26 \\
H13c & 13 & 20.648 & $4.7\times10^{-6}$ & $5.7\times10^{-5}$ & 100.00 & 14.46 \\
H13w & 13 & 18.852 & $3.0\times10^{-5}$ & $9.8\times10^{-5}$ & 100.00 & 23.84 \\
H15c & 15 & 21.973 & $8.9\times10^{-7}$ & $1.9\times10^{-5}$ & 100.00 & 22.85 \\
H15w & 15 & 21.681 & $1.0\times10^{-5}$ & $6.0\times10^{-5}$ & 100.00 & 40.05 \\
\end{tabular}
\end{ruledtabular}
\caption{Hadamard-primitive performance at common peak nominal Rabi frequency. $\mathcal R_A=\Phi/\theta$ uses $\theta=\pi/2$, and $N$ counts the constituent rotations in a given sequence. Means, sampled maxima and $f_4(0.15)$ refer to $B_{0.15}$; $f_4^{(0)}(0.5)$ is the fraction of $B_{0.5}$ occupied by a connected region with $\mathcal I\leq10^{-4}$. The two fractions use different, explicitly specified domains. Means and fractions use $801\times801$ midpoint grids, with boundaries additionally checked for maxima. All entries use the full-precision controls. The AP entries are the published branches with the lowest independently evaluated mean at each amplitude order. }
\label{table:H_comparison}
\end{table*}

\section{Performance and resource trade-offs}
\label{Sec:objective_selection}
\label{Sec:nominal_tradeoffs}

The reference sequences in
Tables~\ref{table:resource_comparison} and
\ref{table:H_comparison} span complementary approaches
to compensating systematic errors. BB1 and SK1 correct amplitude errors, whereas CORPSE corrects detuning errors; concatenation combines the two mechanisms for arbitrary transverse rotations~\cite{NRPBCP,Brown_2004,Cummins_2003,Bando_2013}. The reduced CORPSE-in-SK1 and CORPSE-in-BB1 sequences, denoted rCinSK and rCinBB, remove redundant rotations from these constructions. Reduced SKinsC reverses the nesting order and provides a lower-area alternative. SCORBUTUS provides an alternative based on
a switchback modification of SCROFULOUS, compensating
first-order amplitude and detuning errors for arbitrary
transverse rotations~\cite{Kukita_2022_switchback}. These arbitrary-angle references apply to both targets considered here. Universal inversion sequences provide a different starting point: their phases are aligned with the nominal $X$ axis before the full gate infidelity is evaluated~\cite{PhysRevLett.113.043001,Genov_2020_profile}.

For the $X$ target, the seven-pulse Jones constructions, the symmetric nine-pulse Jones S9 sequence and the balanced ASBO-9 branches provide compact analytical NOT-gate references~\cite{Jones_2013_NOT,Odedra_2012}. Their constituent $\pi$ pulses give direct comparisons at areas $7\pi$ and $9\pi$. The higher-order CORPSE-completed AP/PD references~\cite{arbitraryCp} extend amplitude-error cancellation at larger area; they remain included in Tables~\ref{table:resource_comparison} and \ref{table:H_comparison}. Recursive constructions also reach higher orders: $F_2$ contains 25 $\pi$ pulses, while a nested ASBO-9 construction contains 81~\cite{Jones_2013_nested}. Appendix~\ref{app:reference_controls} specifies the phase conventions and reference constructions used in these comparisons.

\begin{figure*}[tp]
\includegraphics[width=0.99\textwidth]{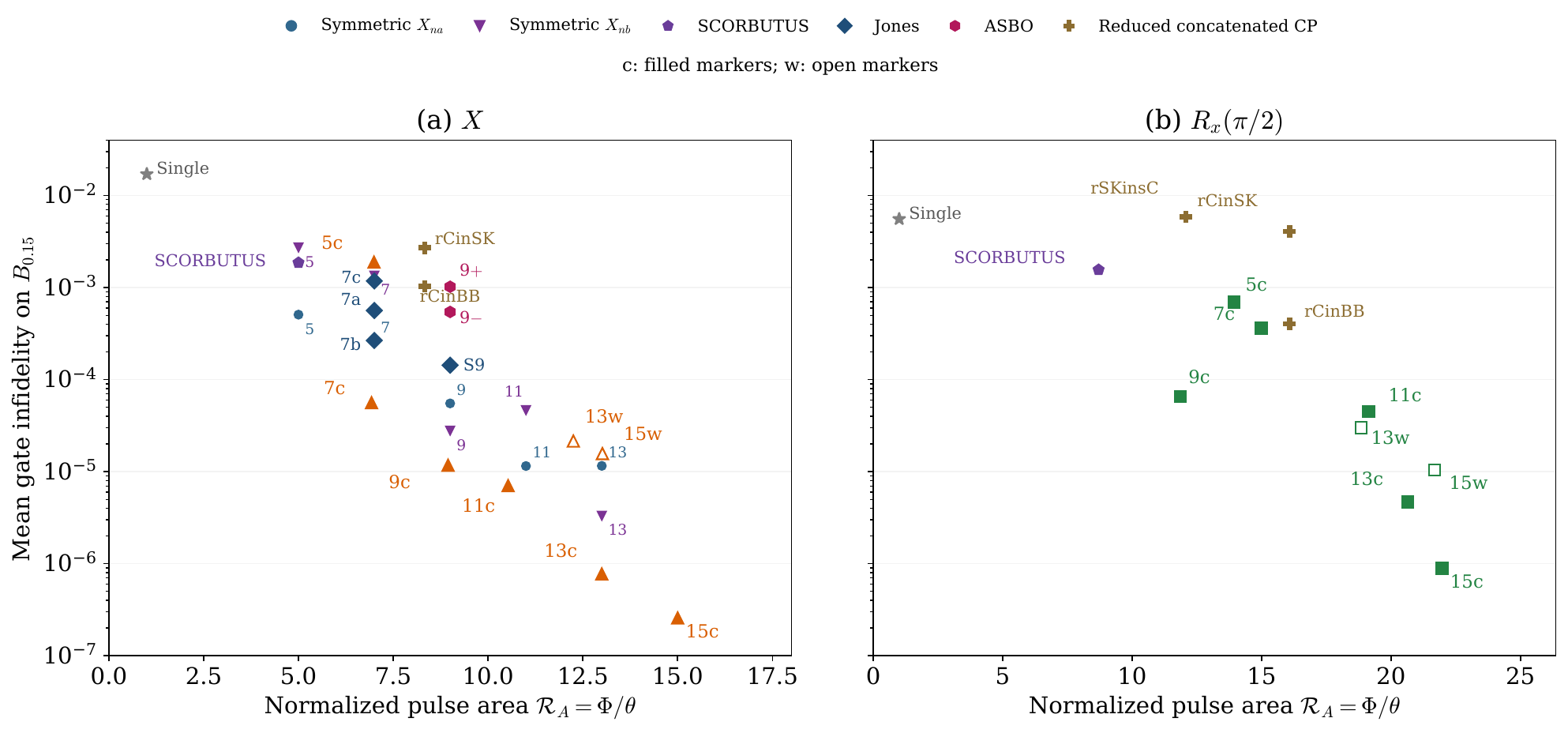}
\caption{Mean infidelity on $B_{0.15}$ versus normalized pulse area for (a) $X$ and (b) $R_x(\pi/2)$. Circles and inverted triangles denote the symmetric $X_{na}$ and $X_{nb}$ families, with numbers identifying $N$. Orange triangles and green squares identify numerical $X$ and Hadamard-primitive controls; filled markers denote c sequences and open markers denote the paired width-oriented w sequences. Diamonds, hexagons and crosses distinguish Jones, ASBO and reduced concatenated sequences, respectively; purple pentagons denote SCORBUTUS. Their labels identify members of the corresponding legend families. Tables~\ref{table:resource_comparison} and \ref{table:H_comparison} give the full comparison, including higher-area references.}
\label{fig:normalized_resources}
\end{figure*}

Figure~\ref{fig:normalized_resources}(a) shows the approximately exponential improvement in central accuracy along the investigated symmetric $X_n$ families. Since every constituent pulse has area $\pi$, $\mathcal R_A=N$, and the overall decrease appears as an approximately linear trend on the semilogarithmic axes. Their mean infidelities fall from $5.08\times10^{-4}$ and $2.74\times10^{-3}$ at $N=5$ to $1.16\times10^{-5}$ and $3.33\times10^{-6}$ at $N=13$ for the a and b branches, respectively. The variable-area precision designs extend the accessible accuracy: X13c and X15c reach means $7.74\times10^{-7}$ and $2.57\times10^{-7}$ at areas $13\pi$ and $15\pi$. Thus the added control length supports a substantial gain in accuracy within the prescribed error window.

The compact analytical references place this gain in context. At seven pulses, X7a and Jones 7b have nearly equal means, while Jones 7c has a lower mean than X7b on $B_{0.15}$; the shorter solutions therefore also remain competitive. At nine pulses, our CP produce a clearer separation: X9a and X9b reduce the Jones S9 mean by factors of approximately 2.6 and 5.1 at identical pulse count, area and duration. X9b also has a mean about 19 times below that of ASBO-9$-$ on this domain. At eleven pulses, X11a has mean $1.15\times10^{-5}$, compared with $1.32\times10^{-3}$ and $2.13\times10^{-4}$ for U11a and U11b at the same area $11\pi$. The numerical X11c reaches $7.05\times10^{-6}$ with the smaller area $10.529\pi$.

The Hadamard results in Fig.~\ref{fig:normalized_resources}(b) show a similar progression toward lower mean error as the longer numerical controls become available. Their comparison with concatenated rotations also exposes the cost of that improvement. H5c reduces the rCinSK mean from $4.08\times10^{-3}$ to $6.93\times10^{-4}$ at lower area, but rCinBB remains more accurate than H5c. H7c improves on the rCinBB mean by about $10.7\%$ while using $6.8\%$ less area and one additional segment. H15c reaches $8.9\times10^{-7}$, approximately 454 times below rCinBB, with only about $37\%$ greater area. The progression therefore includes both gains at comparable area and larger gains obtained with additional resources.

Central accuracy and tolerance of larger errors are complementary outcomes of the numerical construction. In Fig.~\ref{fig:normalized_resources}, the c designs lie below their w counterparts because they minimize the mean error in the central window. As already noted for the 15-pulse $X$ pair, the connected accurate fraction of $B_{0.5}$ increases from $21.30\%$ to $43.57\%$, while the central mean changes from $2.57\times10^{-7}$ to $1.57\times10^{-5}$. The Hadamard pair increases the fraction from $22.85\%$ to $40.05\%$, with central means $8.9\times10^{-7}$ and $1.0\times10^{-5}$. Both paired designs therefore offer an explicit choice between deeper suppression in a calibrated window and broader tolerance of amplitude and detuning offsets, using the same number of pulses.

\section{Summary and Conclusion}
\label{Sec:conclusions}

We have constructed composite $X$ and Hadamard gates that suppress simultaneous quasistatic amplitude and detuning errors in a driven two-level system. The construction combines cancellation of derivatives of the full propagator with numerical design over finite error domains. At five pulses, the derivative conditions recover known universal phase families; at seven pulses, they select members of Jones’s symmetric family, with longer pulses outperforming all benchmarked references in the comparisons and error domains examined here. Moreover both derivative cancellation and numerical designs exhibit an approximately exponential decrease in central mean infidelity over the derived CPs. 

Allowing the pulse areas to vary provides further control over the error landscape. The numerical $X$ and Hadamard catalogues extend to fifteen pulses and include paired central-accuracy and width-oriented designs at thirteen and fifteen pulses. Within the paired resource bounds, these designs favor either deeper suppression in the central window or a larger connected region below infidelity $10^{-4}$. For example, the fifteen-pulse $X$ pair increases the connected fraction of the wider error square from \(21.30\%\) to \(43.57\%\), at the expense of a higher central mean infidelity. We thus show that local cancellation order, mean error, and threshold coverage distinguish different properties of a composite gate.

The phase-precision calculations show that the deepest error suppression places greater demands on phase control. Tests with finite pulse edges and continuous detuning fluctuations further characterize selected fixed sequences, with a response that depends on both the control and the noise spectrum.

The present calculations describe coherent two-level dynamics and assume ideal virtual $Z$ rotations. Natural next steps include extending these constructions to entangling two-qubit gates, incorporating decoherence, cross-talk model, and hardware calibrated cost functions. Together, these directions can further close the gap to fault-tolerant operation while preserving the hardware friendliness of composite control.

\section*{Acknowledgments}
This research is supported by the Bulgarian national plan for recovery and resilience, Contract No. BG-RRP-2.004-0008-C01 (SUMMIT), Project No. 3.1.4, and by the European Union's Horizon Europe research and innovation program under Grant Agreement No. 101046968 (BRISQ).

\appendix
\section{Precision of the control phases}
\label{app:phase_precision}

Errors in the relative pulse phases perturb the interference
responsible for composite-pulse error cancellation. We examine
their effect on X9a, H9c, X15c, and H15w, covering both
constant-area and variable-area controls. Two effects are considered separately:
independent random phase offsets and deterministic rounding to a finite digital phase. Pulse amplitudes and durations remain fixed throughout. All calculations use rectangular
pulses, and the virtual $Z$ rotations completing the Hadamard gates remain ideal.

\subsection{Random phase offsets}

We assign an independent, zero-mean Gaussian offset to each
pulse phase,
\begin{equation}
\phi_k^{\rm actual}=\phi_k+\eta_k,
\qquad
\eta_k\sim\mathcal N(0,\sigma_\phi^2),
\end{equation}
where each offset is constant during its pulse and
$\sigma_\phi$ is the RMS phase error in radians. At each fixed amplitude--detuning error pair, we average the gate infidelity over these offsets, denoting the result by $\langle\mathcal I\rangle_\phi$.

The nominal phase sensitivity has a simple form for an
$X$ gate composed of an odd number $N$ of resonant $\pi$
pulses. The phase offsets enter the resulting rotation
through the alternating sum
\begin{equation}
\Delta\phi=\sum_{k=1}^{N}(-1)^{k+1}\eta_k,
\qquad
\langle(\Delta\phi)^2\rangle=N\sigma_\phi^2.
\end{equation}
Since the gate infidelity is
$\mathcal I=\tfrac23\sin^2(\Delta\phi)$, Gaussian averaging
gives
\begin{equation}
\left.\langle\mathcal I\rangle_\phi\right|_{\epsilon=\delta_c=0}
=\frac{1-e^{-2N\sigma_\phi^2}}{3}
\simeq\frac{2N}{3}\sigma_\phi^2,
\qquad N\sigma_\phi^2\ll1.
\label{eq:phase_floor}
\end{equation}
For this error model, the nominal phase-error contribution
is therefore fixed by $N$ and $\sigma_\phi$, even when the
sequence cancels amplitude and detuning errors. Equation~
\eqref{eq:phase_floor} applies directly to X9a. The
variable-area controls are evaluated using their full pulse propagators.

Figure~\ref{fig:phase_precision}(a) shows the quadratic
dependence on small phase offsets. At
$\sigma_\phi=10^{-4}$ rad, the nominal phase-averaged
infidelities range from $4.36\times10^{-8}$ to
$9.69\times10^{-8}$. Their effect on the useful accuracy of a sequence depends on how these contributions compare with its error in the absence of phase offsets.

To assess this effect away from the nominal point, we
additionally average $\langle\mathcal I\rangle_\phi$ over
$B_{0.15}$ using a $201\times201$ midpoint grid
in Fig.~\ref{fig:phase_precision}(b). At
$\sigma_\phi=10^{-4}$ rad, the domain means increase by
less than $1\%$ for X9a, H9c, and H15w. For X15c, however,
the mean rises from $2.57\times10^{-7}$ to
$3.51\times10^{-7}$, an increase of approximately $37\%$.
The absolute increase, about $9.4\times10^{-8}$, is
comparable to those of the other controls; its larger
relative effect follows from the much smaller X15c
baseline. At $\sigma_\phi=10^{-3}$ rad, the X15c mean
reaches $9.67\times10^{-6}$. Retaining the benefit of
deep central suppression therefore requires the phase-error contribution to remain small compared with the original domain mean.

\subsection{Finite digital phase resolution}

To isolate digital rounding, we set the random phase offsets to zero and round each requested phase to the nearest value on a $b$-bit grid,
\begin{equation}
\phi_k^{(b)}
=q_\phi\,\operatorname{round}\!\left(\frac{\phi_k}{q_\phi}\right),
\qquad
q_\phi=\frac{2\pi}{2^b}.
\end{equation}
Each rounding error is bounded by $q_\phi/2$. 

\begin{figure*}[tp]
\includegraphics[width=0.99\textwidth]{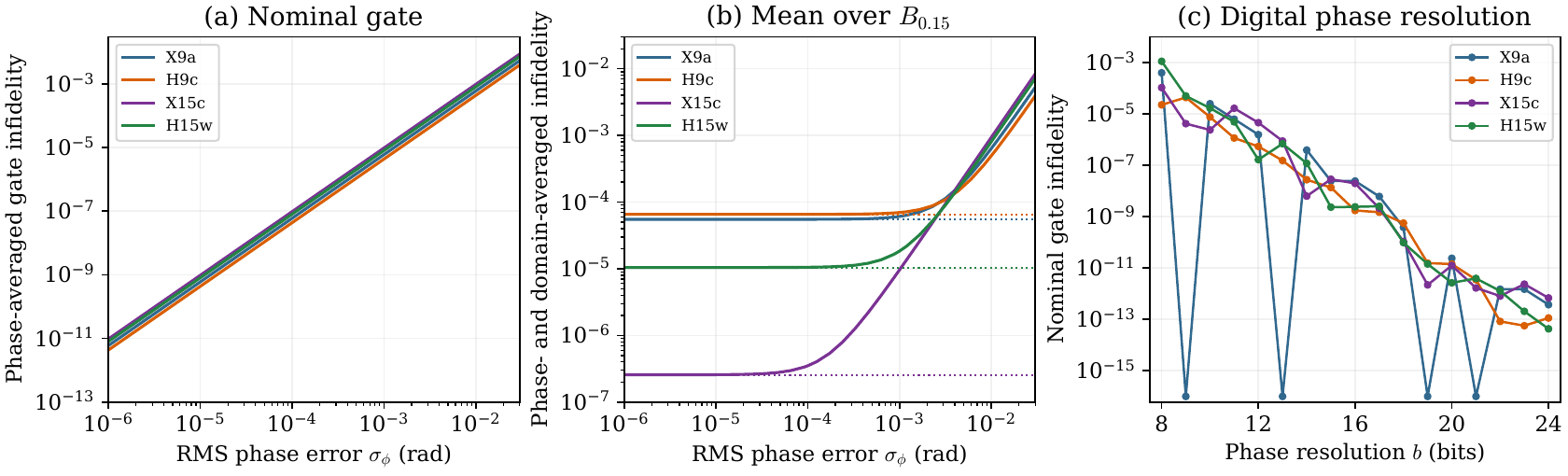}
\caption{Phase sensitivity of X9a, H9c, X15c, and H15w.
Independent Gaussian pulse-phase offsets of RMS magnitude
$\sigma_\phi$ give the phase-averaged gate infidelity
(a) at $\epsilon=\delta_c=0$ and (b) additionally averaged
over $B_{0.15}$. Dotted lines in (b) indicate the domain
means without phase offsets. (c) Nominal gate infidelity
after deterministic rounding of the phases to a $b$-bit
grid, with no additional random offsets.}
\label{fig:phase_precision}
\end{figure*}

Figure~\ref{fig:phase_precision}(c) shows the resulting
nominal infidelities. Increasing $b$ reduces the bound on
each individual phase error, but the gate infidelity need
not decrease at every step. Rounding errors that cancel
at one resolution can reinforce at another. For X9a,
this follows directly from the alternating phase sum
above.

At 16-bit resolution, the phase spacing is
$q_\phi=9.59\times10^{-5}$ rad, and all four nominal
infidelities are below $2.46\times10^{-8}$. Over
$B_{0.15}$, the absolute relative change in the mean is
below $0.4\%$ for X9a, H9c, and H15w, while the X15c
mean increases by approximately $8.0\%$. Thus small
nominal rounding errors can still produce an appreciable
relative change in a deeply suppressed domain mean.

The control coefficients, phase-averaging and rounding
calculations, and grid-convergence checks are provided
in the reproduction repository~\cite{my_github_repo}.

\section{Finite pulse edges and time-dependent detuning}
\label{app:pulse_shape}

We examine how finite switching times and detuning fluctuations affect
X9a, X13c, X13w, H9c, H13c, and H13w. All control coefficients remain
fixed. To compare the sequences at a common peak Rabi frequency
$\Omega_{\max}/(2\pi)=5$ MHz, we set
\begin{equation}
\Omega_k=\frac{a_k}{a_{\max}}\Omega_{\max},
\qquad
T_s=\frac{\pi a_{\max}}{\Omega_{\max}},
\end{equation}
where $a_{\max}=\max_k a_k$ and $T_s$ is the rectangular segment
duration. This rescaling preserves the pulse areas,
$\Omega_kT_s=\pi a_k$. The virtual $Z$ rotations completing the
Hadamard gates remain ideal.

\begin{figure*}[tp]
\includegraphics[width=0.99\textwidth]{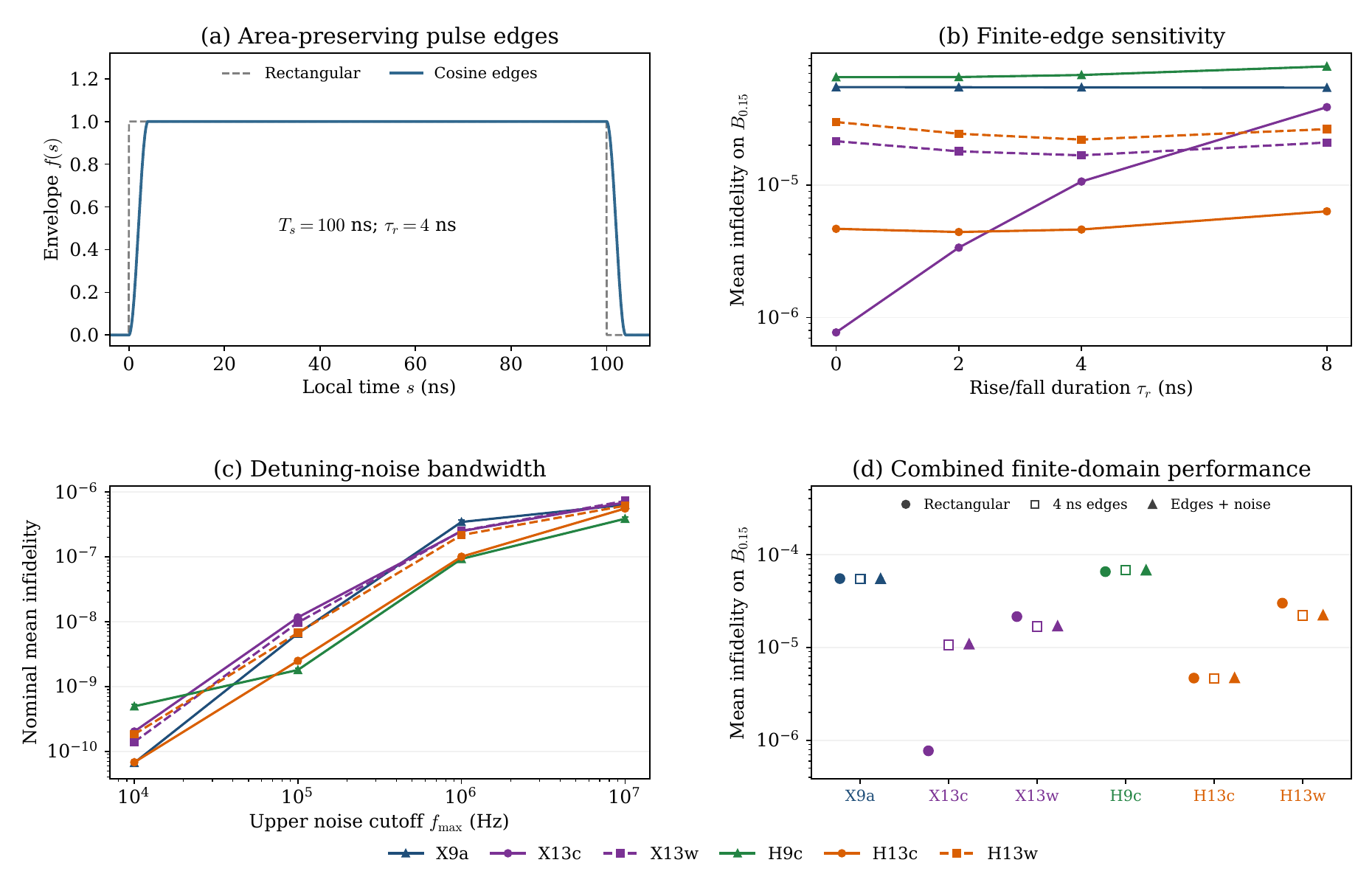}
\caption{Finite-edge and detuning-noise sensitivity of three $X$ and
three Hadamard composite gates.
(a) Area-preserving envelope illustrated for $T_s=100$ ns and
$\tau_r=4$ ns.
(b) Mean infidelity over $B_{0.15}$ versus edge duration, without
time-dependent detuning noise.
(c) Noise-averaged infidelity at $\epsilon=\delta_c=0$ versus the upper
spectral cutoff, with 4 ns edges. The integrated noise power increases
with the cutoff.
(d) Mean infidelities over the same sampled square for rectangular
pulses, 4 ns edges, and 4 ns edges with detuning noise in the
1 Hz--1 MHz band. Error bars in (c,d) show the standard error of the
noise-ensemble mean.}
\label{fig:pulse_shape_sensitivity}
\end{figure*}

We replace each rectangular segment by $\Omega_k f(s)$, where $s$
is the time from the start of the segment and
\begin{equation}
f(s)=
\begin{cases}
\tfrac12[1-\cos(\pi s/\tau_r)],&0\leq s<\tau_r,\\
1,&\tau_r\leq s\leq T_s,\\
\tfrac12[1+\cos(\pi(s-T_s)/\tau_r)],&
T_s<s\leq T_s+\tau_r,\\
0,&\text{otherwise}.
\end{cases}
\label{eq:flat_top}
\end{equation}
For $0<\tau_r<T_s$, this envelope has integral $T_s$ and duration
$T_s+\tau_r$ [Fig.~\ref{fig:pulse_shape_sensitivity}(a)]; we treat
$\tau_r=0$ as the rectangular limit. The pulse areas and peak drive
are unchanged, but the duration of an $N$-pulse sequence increases
by $N\tau_r$.

Without detuning, the Hamiltonian of a fixed-phase pulse commutes with
itself at different times, so its propagator depends only on the pulse
area. The envelope therefore preserves the rectangular-pulse rotation,
also for a constant amplitude error. Static or time-dependent detuning
breaks this commutation property. The resulting propagator depends on
the envelope and the additional time spent under detuning.

We model time-dependent detuning as
\begin{equation}
\Delta(t)=\Omega_{\max}\delta_c+2\pi\xi_f(t),
\qquad S_{\xi}(f)=\frac{A}{f},
\label{eq:detuning_noise}
\end{equation}
where $\xi_f(t)$ is a stationary, zero-mean Gaussian frequency
fluctuation in Hz, continuous across the complete sequence. The
one-sided power spectral density $S_{\xi}(f)$ is nonzero only for
$f_{\min}\leq f\leq f_{\max}$. Motivated by measurements of
low-frequency qubit-frequency fluctuations~\cite{Yan_2012_noise,Burnett_2019_frequency},
we use $A=3.6\times10^5\,{\rm Hz}^2$ and $f_{\min}=1$ Hz. The
frequency variance is
\begin{equation}
\sigma_f^2=A\ln\!\left(\frac{f_{\max}}{f_{\min}}\right).
\end{equation}
Our principal comparison uses $f_{\max}=1$ MHz, giving an RMS frequency
fluctuation of $2.23$ kHz. In the cutoff sweep, $A$ and $f_{\min}$
remain fixed, so increasing $f_{\max}$ increases both the bandwidth
and the total noise power.

We calculate domain means on an $81\times81$ midpoint grid over
$B_{0.15}$ in $(\epsilon,\delta_c)$. Noise averages use 64 independent
antithetic pairs for the domain means and 256 pairs at each cutoff
in the nominal-point sweep. Each pair contains a trajectory and its
negative. Standard errors are calculated from
the independent pair averages.

Finite edges affect the sequences differently
Fig.~\ref{fig:pulse_shape_sensitivity}(b). With 4 ns edges, the
X9a mean remains close to $5.5\times10^{-5}$, whereas the X13c mean
increases from $7.7\times10^{-7}$ to $1.1\times10^{-5}$.
For X13w, the mean instead decreases from $2.2\times10^{-5}$ to
$1.7\times10^{-5}$. The deeper suppression achieved by X13c for
rectangular pulses is more sensitive to this envelope change, although
its mean remains below that of X13w. Preserving pulse areas therefore
does not preserve the error landscape, and finite edges can either
raise or lower the domain mean.

At $\epsilon=\delta_c=0$, all six sequences show increasing
noise-averaged infidelity as $f_{\max}$ increases
Fig.~\ref{fig:pulse_shape_sensitivity}(c). At this point, the
area-preserving envelope leaves the noise-free rotations unchanged.
The sweep thus measures the response to detuning fluctuations, but
does not isolate their frequency dependence because the total noise
power also increases.

Over $B_{0.15}$, adding noise in the principal band increases the
finite-edge means by approximately $0.1\%$--$2.1\%$
Fig.~\ref{fig:pulse_shape_sensitivity}(d). For H13c, the means are
$4.7\times10^{-6}$ for rectangular pulses, $4.6\times10^{-6}$ with
4 ns edges, and $4.7\times10^{-6}$ with edges and noise. For H13w,
the corresponding values are $3.0\times10^{-5}$, $2.2\times10^{-5}$,
and $2.2\times10^{-5}$. At this noise strength, the envelope change
accounts for the largest shifts in the domain means. The nominal-point
results resolve the noise contribution without the larger background
infidelity away from the origin.

The three Hadamard controls were designed using rectangular and
4 ns finite-edge models through Eq.~\eqref{eq:H_selection} and the
corresponding threshold objective. Thus these results 
test performance within that model. The $X$ controls were designed purely for rectangular pulses, so these calculations test their sensitivity to envelope changes and detuning fluctuations absent from their
design. The reproduction materials provide the sampling details and
numerical convergence checks~\cite{my_github_repo}.

\section{Numerical controls and reference constructions}
\label{app:numerical_controls}

Tables~\ref{table: paramsX} and \ref{table: params} list the numerical
$X$ and Hadamard-primitive controls. The total pulse area is
$\Phi=\sum_k\Omega_kT$. Full-precision controls, optimization records,
and reproduction code are available in the repository~\cite{my_github_repo}.
The following subsections specify the reference constructions and
their conversion to the phase convention used here.

\begin{table*}[tp]
\centering\small
\begin{ruledtabular}
\begin{tabular}{llr}
Control & $(\Omega_1T,\ldots,\Omega_NT)/\pi$; $(\phi_1,\ldots,\phi_N)/\pi$ & $\Phi/\pi$ \\
X5c & $(0.99572,2.00000,1.00105,2.00000,0.99572)$ & 6.992 \\
 & $(0.65639,0.96418,0.30954,0.96418,0.65639)$ &  \\
X7c & $(0.98733,0.84159,1.14120,0.98586,1.14120,0.84159,0.98733)$ & 6.926 \\
 & $(0.38978,0.14290,0.13222,0.75938,0.13222,0.14290,0.38978)$ &  \\
X9c & $(0.99182,0.97414,0.89180,1.11405,0.98708,0.98913,1.11120,0.77126,$ & 8.945 \\
 & $1.11405)$ &  \\
 & $(1.39467,0.78094,0.33846,0.25097,0.75708,1.79069,1.36821,1.43204,$ &  \\
 & $1.39572)$ &  \\
X11c & $(0.96052,0.95720,0.96127,0.96367,0.88999,1.01511,0.95762,0.95741,$ & 10.529 \\
 & $0.95077,0.95293,0.96202)$ &  \\
 & $(0.04201,0.99033,1.87099,1.05409,0.66840,0.69101,1.19305,0.16938,$ &  \\
 & $1.57679,1.24324,0.79641)$ &  \\
X13c & $(1.07688,0.99101,0.95544,1.01220,0.97107,0.91350,1.13136,0.95325,$ & 13.000 \\
 & $0.96548,1.07088,0.79685,1.10012,1.06196)$ &  \\
 & $(1.02088,1.95036,1.59783,1.79813,1.03237,1.68959,1.94393,1.56179,$ &  \\
 & $0.76804,1.41803,1.11533,1.22554,0.16717)$ &  \\
X13w & $(1.04620,0.95549,0.83389,0.97612,0.89991,0.85087,1.12508,0.85087,$ & 12.250 \\
 & $0.89991,0.97612,0.83389,0.95549,1.04620)$ &  \\
 & $(0.64959,1.65668,1.39736,1.70874,1.01635,1.72133,0.07511,1.72133,$ &  \\
 & $1.01635,1.70874,1.39736,1.65668,0.64959)$ &  \\
X15c & $(1.14030,0.71719,1.07581,1.02455,1.07993,0.74806,1.14030,1.00143,$ & 15.000 \\
 & $1.01236,1.02072,0.97860,1.00006,0.97918,1.05220,1.02933)$ &  \\
 & $(0.79516,0.87911,0.59835,1.83680,0.59310,0.86406,0.71357,0.33411,$ &  \\
 & $1.76825,0.54515,0.52899,1.79431,0.38332,0.04965,0.89549)$ &  \\
X15w & $(0.87036,0.87403,0.87403,0.86540,0.85848,0.86182,0.86418,0.87259,$ & 13.016 \\
 & $0.87234,0.87189,0.86993,0.85932,0.85419,0.87403,0.87321)$ &  \\
 & $(1.15452,1.99994,0.84443,1.92387,1.25995,0.80723,0.61339,0.71220,$ &  \\
 & $1.14563,1.96763,1.18826,0.74101,0.48888,0.21606,1.67308)$ &  \\
\end{tabular}
\end{ruledtabular}
\caption{Numerical $X$ controls, with pulse areas and phases in units of $\pi$, rounded to five decimal places. For each control, the area vector precedes the phase vector; both follow chronological order. All segments have duration $T$, with $\Omega_0T=\pi$. The last column gives the total pulse area. At $N=13,15$, the suffixes $c$ and $w$ distinguish the paired central-accuracy and width objectives.}
\label{table: paramsX}
\end{table*}

\begin{table*}[tp]
\centering\small
\begin{ruledtabular}
\begin{tabular}{llr}
Control & $(\Omega_1T,\ldots,\Omega_NT)/\pi$; $(\phi_1,\ldots,\phi_N)/\pi$ & $\Phi/\pi$ \\
H5c & $(1.19019,1.83575,0.91729,1.83575,1.19019)$ & 6.969 \\
 & $(1.23357,0.19706,1.39028,0.19706,1.23357)$ &  \\
H7c & $(1.25493,0.86904,1.14647,1.01548,0.69437,1.25493,1.25493)$ & 7.490 \\
 & $(1.25669,0.21358,0.27208,1.42762,0.16186,0.17848,1.20104)$ &  \\
H9c & $(0.66411,0.64293,0.70458,0.67366,0.81848,0.81848,0.65075,0.65315,$ & 5.929 \\
 & $0.30328)$ &  \\
 & $(0.60695,0.46657,0.64915,0.97025,1.86085,1.41624,1.45469,1.77454,$ &  \\
 & $0.29643)$ &  \\
H11c & $(0.94470,1.15353,0.91924,0.67587,0.85261,1.15353,0.44461,0.69395,$ & 9.571 \\
 & $0.94334,0.68117,1.10830)$ &  \\
 & $(0.82739,0.76270,0.15796,0.25168,0.83091,0.12632,0.24442,0.22091,$ &  \\
 & $0.90257,0.77882,0.15808)$ &  \\
H13c & $(0.82072,0.65339,0.56647,0.89759,0.71521,0.90756,0.96755,0.96755,$ & 10.324 \\
 & $0.56295,0.95563,0.76426,0.84073,0.70459)$ &  \\
 & $(0.43564,0.72841,0.61883,0.05985,0.08742,0.70097,0.02200,0.06269,$ &  \\
 & $0.02995,0.65062,0.60437,1.92092,0.36117)$ &  \\
H13w & $(0.63477,0.54508,0.73648,0.82586,0.80326,0.82002,0.87078,0.74412,$ & 9.426 \\
 & $0.87078,0.78940,0.87078,0.53777,0.37675)$ &  \\
 & $(0.86891,0.90819,0.68008,1.97714,0.11468,0.81464,0.09143,0.04642,$ &  \\
 & $0.21566,1.00567,0.90276,0.13198,0.35736)$ &  \\
H15c & $(0.58134,0.56535,0.82745,0.82046,0.72780,0.77823,0.71309,0.84092,$ & 10.987 \\
 & $0.84092,0.71782,0.81599,0.81937,0.73565,0.58057,0.62163)$ &  \\
 & $(0.32805,0.40400,1.48496,1.80162,1.74390,0.96997,1.62019,1.81859,$ &  \\
 & $1.47500,0.51353,0.78681,0.81744,0.02164,0.58771,0.65885)$ &  \\
H15w & $(0.60940,0.52512,0.76526,0.77207,0.78475,0.75169,0.67222,0.82462,$ & 10.840 \\
 & $0.82462,0.71918,0.79172,0.82462,0.74780,0.61051,0.61679)$ &  \\
 & $(0.31618,0.35529,1.44186,1.75283,1.71987,0.92584,1.56285,1.79165,$ &  \\
 & $1.48275,0.52091,0.75839,0.77736,1.97140,0.56403,0.61484)$ &  \\
\end{tabular}
\end{ruledtabular}
\caption{Numerical controls for the $R_x(\pi/2)$ Hadamard primitive, with pulse areas and phases in units of $\pi$, rounded to five decimal places. For each control, the area vector precedes the phase vector; both follow chronological order. All segments have duration $T$, with $\Omega_0T=\pi$. The last column gives the total pulse area. At $N=13,15$, the suffixes $c$ and $w$ distinguish the paired central-accuracy and width objectives.}
\label{table: params}
\end{table*}

\subsection{Concatenated references}
\label{app:reference_controls}

For the reduced concatenated sequences of Ref.~\cite{Bando_2013},
we use $\theta=\pi$ or $\pi/2$ and define
\begin{equation}
k=\arcsin[\tfrac12\sin(\theta/2)],\qquad
\alpha=\arccos[-\theta/(4\pi)].
\end{equation}
The CORPSE target rotation, in chronological order, is
\begin{align}
C_\theta={}&(2\pi+\theta/2-k)_0(2\pi-2k)_\pi\nonumber\\
&\times(\theta/2-k)_0.
\end{align}
The two sequences are
\begin{align}
\mathrm{rCinSK}:&\quad C_\theta(2\pi)_{-\alpha}(2\pi)_\alpha,\\
\mathrm{rCinBB}:&\quad\pi_\alpha(2\pi)_{3\alpha}\pi_\alpha C_\theta.
\end{align}
All pulses use the same nominal peak amplitude, with their rotation
angles set by their durations. Both sequences have total area
$8\pi+\theta-4k$, corresponding to normalized areas $\Phi/\theta$
of 8.333 for $X$ and 16.080 for $R_x(\pi/2)$. Their full propagators
have vanishing first derivatives with respect to amplitude and
detuning errors.

\subsection{SCORBUTUS}

SCORBUTUS is the five-rotation switchback sequence of
Ref.~\cite{Kukita_2022_switchback}. For the targets $R_x(\theta)$
with $\theta=\pi$ or $\pi/2$, let $u\in(0,\pi]$ solve
\begin{equation}
\frac{\sin u}{u}=\frac{2\cos(\theta/2)}{\pi},
\end{equation}
and define
\begin{align}
\alpha&=\arccos\!\left[-\frac{\pi\cos u}{2u\sin(\theta/2)}\right],\\
\beta&=\alpha-\arccos\!\left[-\frac{\pi}{2u}\right],\\
v&=\arccos\!\left[\frac12\left(1-\frac{\pi\sin^2(u/2)}{u}\right)\right].
\end{align}
We use the principal arccosine branches. In the present convention,
the chronological angles and phases are
\begin{align}
\boldsymbol\theta&=(u,v,\pi+2v,v,u),\\
\boldsymbol\phi&=(-\alpha,-\beta-\pi,-\beta,-\beta-\pi,-\alpha).
\end{align}
Ref.~\cite{Kukita_2022_switchback} uses $\cos\phi\,\sigma_x+\sin\phi\,\sigma_y$;
we negate its phases to match our convention. All five pulses use
the common nominal peak amplitude, with durations
$T_k=\theta_k/\Omega_{\max}$.

For $X$, these expressions give
\begin{align}
\boldsymbol\theta&=\pi(1,1/2,2,1/2,1),\\
\boldsymbol\phi&=\pi(-1/3,-2/3,1/3,-2/3,-1/3).
\end{align}

\subsection{Additional compact references}

For the reduced SKinsC sequence, we use the corrected specification
in Ref.~\cite{Kukita_2022_switchback}. Define
\begin{align*}
k&=\arcsin[\sin(\theta/2)/2],\\
a&=\theta/2-k,\qquad b=2\pi-\theta/2-k,\\
\eta&=\arccos[-(2\pi-\theta)/(4\pi)].
\end{align*}
The chronological angles and phases in the present convention are
\begin{align}
\boldsymbol\theta&=(a,b,2\pi,2\pi,a,a),\\
\boldsymbol\phi&=(0,-\pi,-\pi+\eta,-\pi-\eta,-\pi,0).
\end{align}
All pulses use the common nominal peak amplitude. The full propagator
has vanishing first derivatives with respect to amplitude and detuning
errors.

The Jones seven-pulse references have total area $7\pi$. We use the
exact roots in Eqs.~(56)--(60) of Ref.~\cite{Jones_2013_NOT}
rather than the rounded phases in its table. For Jones 7a,
\begin{align*}
\alpha&=-\arccos[-\sqrt{(4+\sqrt{13})/2}/2],\\
\beta&=\alpha-2\pi/3.
\end{align*}
For Jones 7b and 7c, respectively,
\begin{align*}
\alpha&=-\arccos[(3-\sqrt{61})/16],\\
\alpha&=\arccos[(\sqrt{61}-3)/16].
\end{align*}
We reverse the vectors from Ref.~\cite{Jones_2013_NOT} and negate their phases to obtain the
present chronological convention. Independent high-precision
differentiation verifies first-order amplitude and detuning
cancellation for all three sequences. Jones 7a also cancels both pure
second derivatives but retains a mixed second derivative. Jones 7b
and 7c favor amplitude and detuning tolerance, respectively.

\subsection{Higher-order reference constructions}

We combine the published antipalindromic (AP) and palindromic (PD)
correcting blocks~\cite{arbitraryCp} with a CORPSE target rotation
to form the CinAP and CinPD references. These constructions extend
the comparison to higher-order amplitude-error correction.

For a target angle $\theta$, each correcting block contains $L=2n$
rotations of area $2\pi$. With $\gamma=\theta/(2\pi)$, its phases
satisfy the published conditions
\begin{align}
\Phi_L^j(\boldsymbol\varphi)&=f_L^j(\gamma),
\qquad 1\leq j\leq n,\\
\Phi_L^j(\boldsymbol\varphi)
&=\sum_{h_1<\cdots<h_j}
\exp\!\left[-i\sum_{k=1}^{j}(-1)^k\varphi_{h_k}\right],\\
f_L^j(\gamma)
&=\sum_{k=0}^{j}(-1)^k
\binom{(L+\gamma)/2}{k}
\binom{(L-\gamma)/2}{j-k}.
\end{align}
We reconstruct the published solution branches at high numerical
precision, using their tabulated phases as initial guesses and
retaining the AP or PD phase symmetry.

In chronological order, the complete sequence is
\begin{equation}
C_\theta(2\pi)_{-\varphi_L}\cdots(2\pi)_{-\varphi_1},
\end{equation}
where $C_\theta$ is the CORPSE rotation defined above. At zero
detuning, $C_\theta$ preserves the amplitude-error response of the
target rotation, so the correcting block retains its intended
amplitude-error cancellation order. The CORPSE rotation cancels the
first detuning derivative, and each additional $2\pi$ pulse is
insensitive to detuning to first order.

The reproduction materials~\cite{my_github_repo} contain all reconstructed branches and
their performance metrics. For each target and amplitude-cancellation
order, the main comparison table reports the branch with the lowest
independently evaluated mean infidelity on $B_{0.15}$.

\clearpage
\bibliography{references}
\end{document}